\documentclass[conference, 10pt]{IEEEtran}
\usepackage{graphicx}
\usepackage{listings, multicol}
\usepackage{svg}

\usepackage[utf8]{inputenc}
\usepackage{xcolor}

\definecolor{codegreen}{rgb}{0,0.6,0}
\definecolor{codegray}{rgb}{0.5,0.5,0.5}
\definecolor{codepurple}{rgb}{0.58,0,0.82}
\definecolor{backcolour}{rgb}{0.95,0.95,0.92}

\lstdefinestyle{my style}{
    backgroundcolor=\color{backcolour},   
    commentstyle=\color{codegreen},
    keywordstyle=\color{magenta},
    numberstyle=\tiny\color{codegray},
    stringstyle=\color{codepurple},
    basicstyle=\ttfamily\footnotesize,
    breakatwhitespace=false,         
    breaklines=true,                 
    captionpos=b,                    
    keepspaces=true,                 
    numbers=left,                    
    numbersep=5pt,                  
    showspaces=false,                
    showstringspaces=false,
    showtabs=false,                  
    tabsize=2
}
\usepackage{cite}
\usepackage{booktabs}

\usepackage{caption}
\usepackage{subcaption}

\usepackage{mathtools}
\DeclarePairedDelimiter\ceil{\lceil}{\rceil}
\ifx\pdfoutput\undefined
\usepackage[hypertex,hidelinks,colorlinks=true]{hyperref}
\else
\usepackage[pdftex,hidelinks,colorlinks=true,hypertexnames=false]{hyperref}
\fi
\usepackage{xcolor}
\usepackage{url}

\usepackage{subfiles}
\usepackage{verbatim}
\usepackage{algorithm}
\usepackage{algpseudocode}

\begin{document}

\title{
\LARGE
High-Level FPGA Accelerator Design for \\Structured-Mesh-Based Explicit 
Numerical Solvers\vspace{-10pt}
}
%
%


 \author{
 \authorblockN{Kamalavasan Kamalakkannan, \\Gihan R. Mudalige}
 \authorblockA{Dept. of Computer Science\\University of Warwick, UK\\
 \small\texttt{\{kamalavasan.kamalakkannan}, 
 \\\texttt{g.mudalige\}@warwick.ac.uk\normalsize}\vspace{-20pt}}
 \and
 \authorblockN{Istvan Z. Reguly}
 \authorblockA{Faculty of Information Technology \& Bionics \\ Pazmany Peter 
 Catholic University, Hungary 
 \small\\\texttt{reguly.istvan@itk.ppke.hu}\normalsize}
 \and
 \authorblockN{Suhaib A. Fahmy}
 \authorblockA{King Abdullah University of\\ Science and Technology (KAUST), 
\\Thuwal, Saudi Arabia
 \small\\\texttt{suhaib.fahmy@kaust.edu.sa}\normalsize}
 }

\maketitle
\begin{abstract}
This paper presents a workflow for synthesizing near-optimal FPGA 
implementations for structured-mesh based stencil applications for explicit 
solvers. It leverages key characteristics of the application class, its 
computation-communication pattern, and the architectural capabilities of the FPGA 
to accelerate solvers from the high-performance computing domain. Key new 
features of the workflow are (1) the unification of standard state-of-the-art 
techniques with a number of high-gain optimizations such as batching and spatial 
blocking/tiling, motivated by increasing throughput for real-world work loads 
and (2) the development and use of a predictive analytic model for exploring the 
design space, resource estimates and performance. Three representative 
applications are implemented using the design workflow on a Xilinx Alveo U280 
FPGA, demonstrating near-optimal performance and over 85\% predictive model 
accuracy. These are compared with equivalent highly-optimized implementations of 
the same applications on modern HPC-grade GPUs (Nvidia V100) analyzing time to 
solution, bandwidth and energy consumption. Performance results indicate 
equivalent runtime performance of the FPGA implementations to the V100 GPU, with
over 2$\times$ energy savings, for the largest non-trivial application 
synthesized on the FPGA compared to the best performing GPU-based solution. Our 
investigation shows the considerable challenges in gaining high performance on 
current generation FPGAs compared to traditional architectures. We discuss 
determinants for a given stencil code to be amenable to FPGA implementation, 
providing insights into the feasibility and profitability of a design and its 
resulting performance.
\end{abstract}



\noindent \IEEEoverridecommandlockouts


%
\IEEEpeerreviewmaketitle

\vspace{-5pt}
\section{Introduction}\vspace{-5pt}

\noindent
Field Programmable Gate Arrays (FPGAs) have become highly attractive as 
accelerator architectures by virtue of their high performance, low power 
consumption, and re-programmability. As a result, FPGAs have gained a foothold 
in a wider range of application domains such as cyber 
security~\cite{cousins2016designing}, databases~\cite{owaida2017centaur}, and 
deep learning~\cite{wang2016dlau}. In recent years, the integration of FPGAs 
as first-class accelerator platforms has also attracted significant interest in 
the high-performance (HPC) and scientific computing community, particularly in 
the financial computing domain~\cite{becker2015maxeler}. They have also emerged 
as potential accelerator platforms for cloud 
computing~\cite{fahmy2015virtualized}. However, a key limitation has been the 
design effort needed to produce performant accelerators for FPGAs, 
requiring hardware expertise and an alternative approach to programming that is 
more data-flow oriented. Commercial FPGA vendors have attempted to address this 
problem with high level synthesis (HLS) tools that can translate programs 
written in standard high-level languages such as C/C++ or OpenCL. However, 
these tools still require low level modification of code to produce 
accelerators with optimum performance. 


One solution to this problem is to leverage key characteristics of an 
application, its computation-communication patterns or motifs to explore the 
design space on hardware. Once the best optimization strategy for a 
given motif is identified for the target hardware, it can be used as a design 
template for similar applications, even going so far as to create higher-level 
frameworks such as DSLs that can automatically generate the accelerator 
implementations. Such a strategy has become an important technique in 
developing performance portable massively parallel HPC applications given the 
increasing diversity of processor architectures~\cite{Reguly2017, InPar2012, 
pyfr2016, devito2018}. 

In this paper we apply such an analysis to the domain of structured-mesh-based 
explicit numerical solvers, characterized by stencil computations, targeting 
FPGAs. These codes frequently appear as the core motif in solvers for 
partial differential equation (PDEs). As such, they are used in applications 
from a wide range of fields, including computational fluid dynamics (CFD), 
hydro-dynamics, financial computing, and oil/gas exploration simulations.
The key characteristic is the loop over a “rectangular” multi-dimensional set 
of mesh points using one or more “stencils” to access data. This is in contrast 
to unstructured meshes that require explicit connectivity 
information~\cite{InPar2012} between neighboring mesh elements via mappings.

Considerable previous research has developed a range of strategies to 
synthesize optimized FPGA implementations for stencil codes~\cite{3drtm_2011, 
SDSLc2015, Waidyasooriya2017, soda2018, Zohouri2018, Zohouri2018a, 
Waidyasooriya2019, HeteroCL2019,  Dohi2013, Natale2016, DSLSC_2013}. 
Most recent works utilize HLS tools, usually compiling OpenCL, and target both 
2D and 3D stencil applications. They develop a number of standard techniques, 
ranging from basic methods such as cell-parallel/vectorization, unrolling the 
iterative loop, to more complex transformations such as spatial/temporal 
blocking (tiling), in order to best utilize FPGA resources for maximum 
performance. However, many of these previous works target optimizations 
specific to an application in isolation without developing a design strategy 
that can be applied to other stencil codes. While some~\cite{Waidyasooriya2017, 
Waidyasooriya2019} does attempt to generalize accelerator implementations for 
stencil codes, they only target simpler stencil applications without exploiting 
higher-gain optimizations. A key gap is the lack of a unifying design strategy 
particularly focusing on realistic applications. 

As such, an open question remains regarding how to optimize implementation of 
stencil applications on FPGAs, and how characteristics of the application and 
various optimizations determine performance compared to traditional CPU and GPU 
architectures. To this end, in this paper we present an initial proposal and 
unifying workflow for designing near-optimal FPGA implementations for 
structured-mesh based stencil applications together with analytical models that 
enable exploration of the design space for stencil accelerators on modern FPGAs.
Specifically, we make the following contributions:
\begin{itemize}
\item We propose an implementation \textit{template}, and an accompanying 
step-wise optimization strategy for conversion of structured-mesh, explicit, 
iterative stencil applications to FPGA accelerators. Given the hardware 
resource constraints, we focus on features of the application that are amenable 
for FPGA implementation and optimizations for gaining near-optimal performance. 
A key optimization, novel in this area, is the batched execution of multiple 
independent stencil problems on an FPGA.



\item Targeting current generation Xilinx FPGAs, we present the design and 
optimization of three contrasting, representative explicit stencil solvers, 
comparing a range of alternatives based on resource and performance trade-offs. 
The applications include both 2D and 3D stencil solvers and multiple stencil loops. 

\item We develop a predictive analytic model that provides estimates for 
determining the feasibility of implementing a given stencil application on an 
FPGA using the proposed design strategy. The model calculates the resource 
requirements considering the optimizations implemented, together with memory 
requirements and limitations on operating frequency. It predicts the runtime of 
the resulting FPGA synthesis of the application accurate to within $\pm$15\% 
of the achieved runtime. 



\item Finally, the runtime, bandwidth and power/energy performance of the FPGA 
implementations developed with the proposed strategy are compared with highly 
optimized implementations on a traditional accelerator architecture, 
a modern Nvidia GPU.


\end{itemize}\vspace{-1pt}
\noindent Initial results on current generation Xilinx hardware demonstrate 
competitive performance compared to the best performance achieved for the same 
applications on traditional (in this case GPU) architectures using single 
precision floating-point (SP) representations. To our knowledge, such an 
extended workflow for stencil application development accompanied with a 
predictive model have not been previously presented. We believe that the 
proposed approach provides a promising strategy for use in industrial workloads 
from areas such as financial computing, standardizing the development cycle for 
these platforms.  


The rest of the paper is organized as follows: Section~\ref{sec/background} 
begins with a background on structured-mesh stencil applications together with 
previous work that explored FPGA implementations for this class of applications.
Section~\ref{sec/design} presents our proposed design strategy for 
implementing iterative stencil codes for FPGAs, as a step-by-step methodology, 
starting from the basic stencil loops, down to target FPGA code for the Xilinx 
Alveo U280 accelerator board. Further optimizations for the target synthesis is 
discussed in Section~\ref{sec/opts}. The design including the advanced 
optimizations is then explored through the development of an analytic 
performance model. A performance analysis and benchmarking of the FPGA 
implementations compared against the performance of state-of-the-art optimized 
CPU and GPU implementations is presented in Section~\ref{sec/perf}. Finally, 
conclusions are presented in Section~\ref{sec/conclusions}. \vspace{-0pt}

\vspace{-0pt}
\section{Background}\label{sec/background}\vspace{-5pt}



\noindent The key characteristic of structured-mesh stencil computations is  
loops over a “rectangular” multi-dimensional set of mesh points using one 
or more fixed data access patterns, called \textit{stencils}, to access data. 
The main motivating numerical method here is the solution to Partial 
Differential Equations (PDEs), specifically based on the finite difference 
method. These techniques are used extensively in computational fluid dynamics 
(CFD), computational electro magnetics (CEM) in the form of iterative solvers. 
For example the finite difference scheme for the solution of a generic PDE can 
be given by the 2D explicit equation~(\ref{eq/pde1}):
\begin{equation}\label{eq/pde1}
U^{t+1}_{x,y}= aU^{t}_{x-1,y}+ bU^{t}_{x+1,y}+ cU^{t}_{x,y-1}+ 
dU^{t}_{x,y+1}+eU^{t}_{x,y} 
\end{equation}
\noindent Here, \textit{U} is a 2D mesh and \textit{a, b, c, d,} and \textit{e} 
are coefficients. In this example \textit{U} is accessed at spatial mesh 
points \textit{(x-1,y), (x+1,y), (x,y-1), (x,y+1),} and \textit{(x,y)} which 
forms a five point stencil. In an explicit scheme the computation iterates over 
the full rectangular mesh, updating the solution at each mesh point, for the 
current time step, \textit{t+1}, using the solution from the previous time 
step, \textit{t}. The time step iteration usually continues until a steady 
state solution is achieved. There is a data dependency for the computations 
among multiple time step iterations, but no dependency within the spatial 
iterations. As such, each mesh point calculation within a time iteration can be 
computed in parallel. In contrast, an implicit scheme would update the 
solution at the current time step using values from the same time step, 
further introducing a data dependency within the spatial iterations. This leads to a 
much faster convergence to the final solution, but enforces an order in which a 
computation iterates over the mesh leading to limited parallelism. While both 
explicit and implicit schemes are equally used in production settings in 
scientific computing applications, we focus on explicit iterative solvers in 
this paper given their simplicity and higher parallelism. 

\subsection{Related Work}\vspace{-3pt}

\noindent Many previous works have targeted FPGAs for stencil computations. Early 
works~\cite{3dverilog2009,vhdltemplate2012,SSA2011} used Hardware Description 
Languages (HDL) for describing the architectures. However the 
process required extensive hardware knowledge and a time consuming 
development cycle. The introduction of High-Level Synthesis (HLS) tools has 
significantly improved developer productivity and time to design. As such more 
recent work~\cite{Waidyasooriya2017, Waidyasooriya2019, SDSLc2015, soda2018, 
HeteroCL2019} have all utilized HLS tools for implementing FPGA designs for 
stencil computations. As FPGAs have advanced to incorporate a variety of high 
bandwidth interfaces and memory types, the system level view of an accelerator 
architecture has become more important to achieving overall high performance.

The most comprehensive implementation workflow and optimization methodology 
to date is by Waidyasooriya et. al 
in~\cite{Waidyasooriya2017,Waidyasooriya2019}. The authors use OpenCL and 
propose an optimization strategy for stencil applications targeting Intel 
FPGAs. A number of 2-D and 3-D stencil applications are developed through the 
above strategy, demonstrating up to 950 GFLOPS of achieved computational 
performance on Intel FPGAs. Runtime and bandwidth performance are compared to 
conventional GPU and multi-core CPU implementations. The work, however, limits 
the investigation to applications with only a single stencil loop over the mesh. 
Multiple stencil loops within a single time-step iterative loop are not 
considered. 

A previous implementation of the 3D Reverse Time Migration (RTM) application, 
which has similarities to the RTM application we develop later in this paper, can be found 
in~\cite{3drtm_2011}. The implementation uses early-generation Xilinx FPGAs, 
prior to the introduction of HLS tools, but with designs equivalent to the 
techniques we use in this paper. The work in \cite{Zohouri2018} uses Intel FPGAs 
with a design goal to enable unrestricted input sizes for stencil computations 
without  sacrificing  performance. They combine spatial and temporal blocking to 
avoid input size restrictions, and employ multiple FPGA-specific optimizations 
to tackle the added design complexity. The same authors apply these techniques 
to higher-order stencils in \cite{Zohouri2018a}. The use of spatial and temporal 
blocking is novel, which our work in this paper also addresses, but we extend it 
to variable sized tiling and multi-port implementations, generalizing the 
technique and incorporating it to our overall design workflow. 

A number of previous works have also utilized high-level frameworks for 
generating efficient FPGA accelerators. The SDSLc framework~\cite{SDSLc2015} 
presents the use of source-to-source translation for generating parallel 
executables for a range of hardware platforms. These include CPUs, GPUs 
and FPGAs. The paper details optimizations such as iterative loop unrolling 
and full data reuse within FPGAs. Similarly the SODA framework~\cite{soda2018} 
performs several optimizations including perfect data reuse by minimal reuse 
buffers and data quantization. Additionally it models the performance and 
predicts resource consumption, significantly reducing design time. The authors 
present competitive performance with multi-core CPU implementations and 
state-of-the-art stencil implementations on FPGAs. The main limitations of the 
work are fixed tile size and host based tiling. Due to the DSL's support of only 
declarative programming, it is not clear whether any limitations exists for 
porting of complex kernels using SODA.

The more recent HeteroCL framework~\cite{HeteroCL2019} addresses image 
processing applications. It also supports stencil applications through a SODA 
back-end. The HeteroCL DSL separates algorithm from compute, schedules and 
determines data types, and automatically translates SODA DSL to reflect the 
iteration factor, unroll factor and other parameters such as data width. A deep 
single kernel pipeline generated using the above frameworks usually suffers from 
routing congestion in modern large FPGAs from Xilinx that incorporate  multiple 
Super Logic Regions (SLRs). In \cite{Dohi2013}, Dohi et. al, use the 
proprietary MaxCompiler and MaxGenFD high-level design tools to implement 
finite-difference equations. The work is limited to Maxeler Technologies FPGA 
platforms and does not compare results with other FPGAs, GPUs or CPUs. The 
authors of~\cite{Natale2016} use the polyhedral model and implement a related 
framework to automatically accelerate iterative stencil loops on a multi-FPGA 
system. In contrast~\cite{DSLSC_2013} develop a Scalable Streaming Array to 
implement stencil computations on multiple FPGAs, using a DSL, achieving 
reduced development time and near-peak performance. Automatic code generation 
is also used in~\cite{licht2020stencilflow} and has similarities to our design in this paper. However, it mainly focuses on non-iterative applications with multiple kernels, hence spanning designs over multiple FPGAs. Batching and tiling optimizations are not attempted.

In contrast to the above work, this paper presents a unifying strategy for 
the development of FPGA implementations of both 2-D and 3-D stencil 
applications, including multi-dimensional mesh elements and multiple stencil 
loops. We incorporate many of the optimization techniques in previous works 
that have usually been applied in isolation or on a single application. 
Additionally we introduce a number of further optimizations such as batching to 
achieve higher throughput in real-world/production workloads and settings. We 
also present a predictive analytic model that estimates the feasibility of 
implementing a given stencil application on a given FPGA platform. We compare 
the performance of the FPGA accelerators to equivalent highly-optimized 
implementations of the same applications on modern HPC-grade GPUs analyzing time 
to solution, bandwidth, and energy consumption. To our knowledge, the 3D RTM 
application developed in this work, using HLS tools for the FPGA using our 
design flow,  motivated by real-world stencil codes, is also novel, being one of 
the few non-trivial applications presented in literature. Our investigation 
uncovers the determinants for a given stencil code to benefit from a FPGA 
implementation, providing insights into the feasibility of a design and its 
resulting performance.



\vspace{-0pt}
\section{Accelerator Design}\label{sec/design}\vspace{-5pt}


\noindent FPGAs differs from traditional CPUs and GPUs as they do not present a 
fixed general purpose architecture to be exploited using software. A program, 
made up of a sequence of instructions, is executed on a fixed CPU or GPU 
architecture that does not change. In contrast, an FPGA must be configured with 
an architecture that implements the data flow computation for a specific task. 
This leads to significantly reduced energy 
consumption compared to the CPUs and GPUs, primarily due to the locality of data 
movement within the mapped architecture rather than via multiple reads and 
writes to various stages of the memory hierarchy. The reconfigurability of FPGAs 
offers a significant advantage over the design of custom Application Specific 
Integrated Circuits (ASICs) which is much more time-consuming and costly and 
leads to a fixed architecture that cannot be modified post-design. FPGAs 
comprise a variety of basic circuit elements, including ample look-up-tables 
(LUTs) and registers, large numbers of digital signal processing (DSP) blocks on 
modern devices, block memories (BRAM/URAM), clock modules, and a rich routing 
fabric to connect these elements into a large logical accelerator architecture. 
While these resources are primarily suited to implementation of fixed point 
integer data-paths, they can be used to implement floating point data paths 
too, 
though these typically consume significantly more resources for the same 
computation. In modern Xilinx FPGAs, the overall die is split into a number of
regions called Super Logic Regions (SLR)~\cite{XilinxSSI2012}. Bandwidth within 
an SLR is extremely high (TB/s) due to the wealth of connections and memory 
elements, while between SLRs it is limited by the number of silicon connections 
available.  The BRAMs and URAMs that reside in SLRs provide high-speed, small 
blocks of on-chip memory, typically 10s of MB in total, recent devices like the 
U280 have close coupled High Bandwidth Memory (HBM) (8~GB on the U280) connected 
to multiple SLRs. HLS tools can typically connect multiple on-chip BRAMs or 
URAMs to obtain a larger block of memory. An FPGA board will also include much 
larger, but slower DDR4 (32 GB on the U280) memory as external memory. Managing 
the movement of data between these different types of memory is key to 
achieving high computational performance. The performance of an FPGA 
architecture is hard to predict, as it is impacted by various design 
characteristics beyond the level of parallelization applied. As a design grows 
and begins to occupy a larger portion of the FPGA, routing (i.e. connecting all 
the circuit elements together) becomes more challenging, and can reduce the 
achievable clock frequency and hence overall performance.
\begin{figure}[t]
    \centering
    \includegraphics[width=7.5cm]{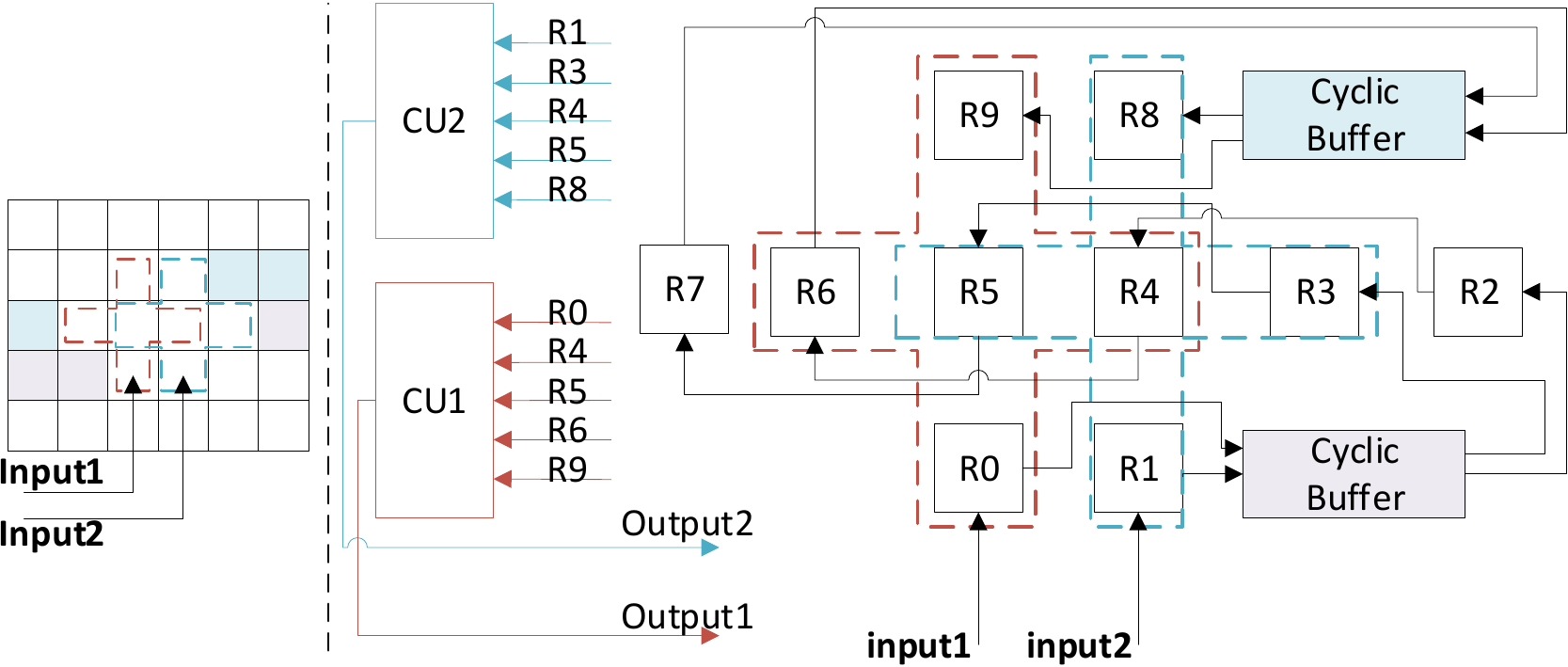}\vspace{-5pt}
    \caption{\small Window buffer and factor of 2 vectorization.}
    \label{fig/vec}\vspace{-20pt}
\end{figure}

To achieve high computational throughput on FPGAs, a custom architecture is 
designed that is then implemented using the low-level circuit elements described 
above. A data-flow arrangement seeks to map a complex computation to a series 
of 
data-paths that implement the required computational steps with movement of 
data 
managed by direct connections. Compared to fixed CPU and GPU architectures where 
the steps in an algorithm are computed sequentially with intermediate results 
stored to registers, FPGA compute pipelines can be much deeper and more 
irregular parallelism can be exploited. Performing a stencil computation will 
then involve, starting up the pipeline (requiring some clock cycles equal to 
the pipeline depth) and outputting the result from the computation for each mesh 
point per clock cycle as a pipelined execution. 

For CPU/GPU architectures such a computation is implemented using nested loops, 
iterating over the mesh and over the neighborhood points. On FPGAs these 
multiple levels of loops can be unrolled. Retaining an outer loop can be costly 
due to the need to flush the unrolled inner loop pipeline which can be long. 
Hence,  multi-dimensional nested loops should be flattened to a 1D loop either 
manually or by using HLS directives such as \texttt{loop\_flatten}. We have 
observed that manual flattening still provides the best performance and 
optimized resource utilization, as current Xilinx HLS compilers can make 
pessimistic scheduling decisions. 


\begin{figure}[t]
    \centering
    \includegraphics[width=8cm]{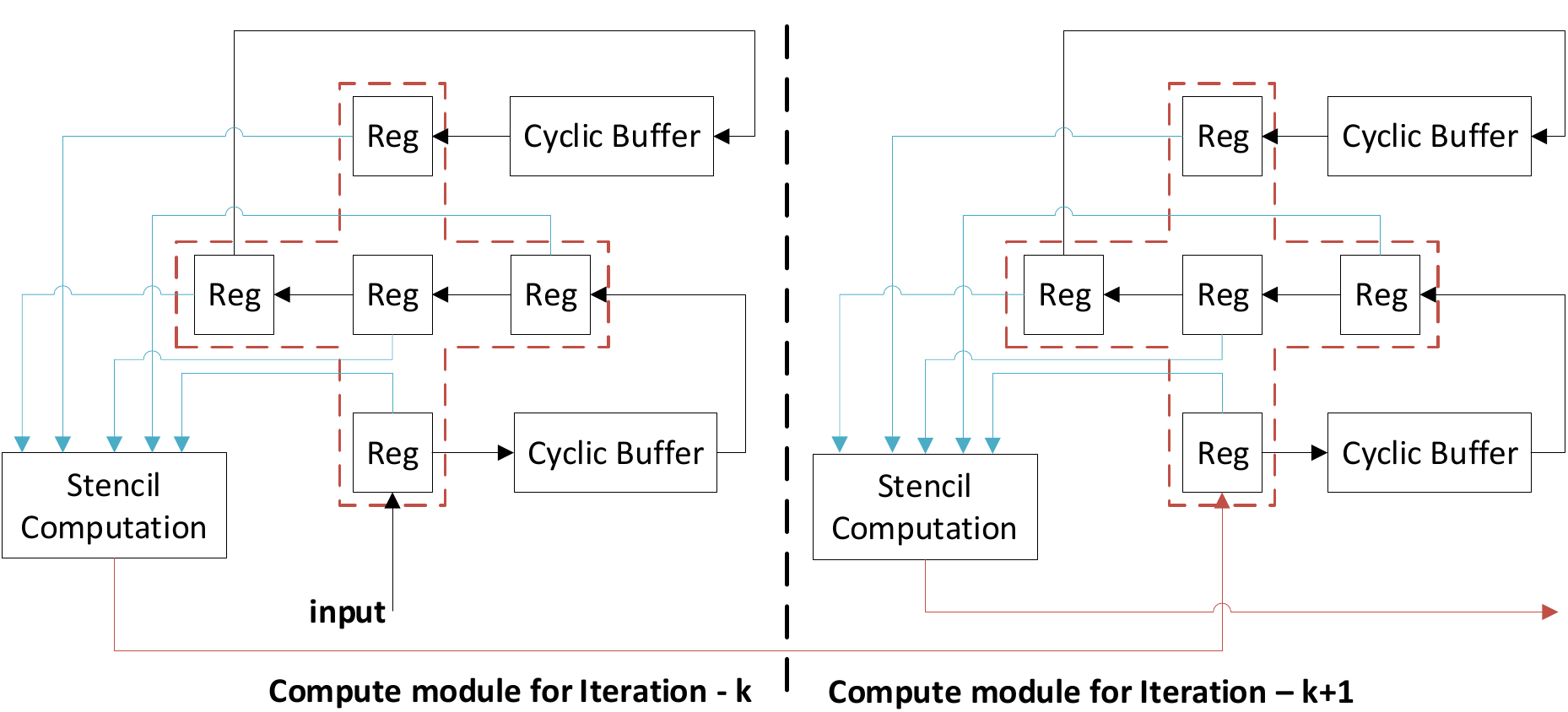}\vspace{-5pt}
    \caption{\small Unrolling the iterative loop.}
    \label{fig/unrolliter}\vspace{-20pt}
\end{figure}

A key approach to gaining the best performance from the above computational 
pipeline is streaming data from/to external and near-chip memories to/from 
on-chip BRAMs/URAMs to feed the computational pipelines efficiently. A perfect 
data reuse path can be created by (1) using a First-In-First-Out (FIFO) buffer 
to fetch data from DDR4/HBM memory without interruption (allowing burst 
transfers) to on-chip memory, and then (2) by caching mesh points using the 
multiple levels of memory, from registers to BRAM/URAM. Fig.~\ref{fig/vec} 
illustrates such a data path for a 2D, 2nd order stencil. This technique has 
previously been referred to as window buffers~\cite{3drtm_2011}. A 2D, $D$ order 
stencil requires $D$ rows to be buffered to achieve perfect data reuse.  
Similarly, $D$ planes should be buffered for a 3D stencil. The total number of 
mesh elements needed to be buffered is the maximum number of mesh elements 
between any two stencil points. BRAM/URAMs can be used to design window buffers 
by using cyclic buffering. Given their high capacity, URAMs are preferred if 
the number of elements to be buffered is large.

Multiple pipelines for the same computation (i.e. loop body or kernel) can be 
created using HLS directives. This technique, called the cell-parallel method 
in~\cite{Waidyasooriya2017} allows computation of the stencil on multiple mesh 
points simultaneously, provided that there are no data dependencies, which is 
the case for explicit schemes. The cell-parallel method is similar to SIMD 
vectorization on CPUs and SIMT on GPUs but on an FPGA it essentially creates 
parallel replicas of the computational units as opposed to single vector 
operations. However the resource availability in an FPGA limits the number of 
parallel units that can be synthesized on a given device. Fig.~\ref{fig/vec} 
illustrates a factor of 2 implementation, where the vectorization factor 
represents the number of mesh points updated in parallel.

Another approach that can increase performance when mapping to FPGAs is to 
unroll the iterative loop, which encompasses one or more stencil loops over the 
rectangular mesh. This allows the results from a previous iteration to be fed to 
the next iteration without writing back to external (DDR4 or HBM) memory. 
This scheme, called the step-parallel technique in previous 
work~\cite{Waidyasooriya2017} is illustrated in Fig.~\ref{fig/unrolliter}. The 
technique leads to increased throughput without the need for additional external 
memory bandwidth. However, the unrolling factor depends once more on available 
FPGA resources and internal memory capacity. Cutting down on external memory 
access in this manner also lead to more power-efficient designs. One 
disadvantage, however, is the increased length of the computational pipeline, 
which significantly affects performance for small mesh sizes. 



\subsection{Model for the Baseline Design}\label{subsec/basemodel}\vspace{-3pt}
\noindent The performance of a baseline design, as discuses above, therefore 
depends on (1) the capacity of the computational pipeline and (2) the external 
memory bandwidth. Computational capacity depends on the number of mesh point 
updates done in parallel (vectorization factor), latency of the pipeline and 
operating clock frequency of the FPGA. However, memory throughput depends on 
various factors such as the number of mesh elements transferred, and the stride 
between each transferred element. To simplify, we model reading/writing of 
contiguous data from/to memory with a maximum transfer size of 4K bytes, to 
reach a near optimal throughput of external/near-chip memory for the Xilinx 
U280 FPGA, our target hardware in this work. 


Assuming that the memory throughput is sufficient to supply $V$ mesh 
points (i.e. a vectorization factor of $V$) continuously without interruption, 
then the total clock cycles taken to process a row from a 2D mesh with $m\times 
n$ elements will be given by $\ceil*{\frac{m}{V}}$. Here, we have padded each 
row to be a multiple of $V$ if required. The compute pipeline will process $n + 
\frac{D}{2}$ rows as there are $D/2$ different rows between the current stencil 
update mesh point and farthest mesh point required for the stencil computation, 
where $D$ is the stencil order. If the outer iterative loop unroll factor is 
given by $p$ then the total number of clock cycles required to process the full 
$m\times n$ mesh for $n_{iter}$ iterations is given by: \vspace{-3pt}
\begin{equation}\label{eq/totclk2D}
Clks_{2D} = \frac{n_{iter}}{p}\times\left(\ceil*{\frac{m}{V}} \times 
(n+p\times\frac {D} {2})\right)
\end{equation}
The above extends naturally to 3D meshes as in (\ref{eq/totclk3D}), where the 
3D mesh size is given by $m\times n\times l$ and $D$ is then equivalent to the 
number of plains to be buffered.\vspace{-3pt}
\begin{equation}\label{eq/totclk3D}
Clks_{3D} = \frac{n_{iter}}{p}\times\left(\ceil*{\frac{m}{V}} \times n \times 
(l+p \times {\frac {D} {2}})\right)\vspace{-3pt}
\end{equation}
As noted before, the models above only hold for cases where the vectorization 
factor $V$, which determines the number of parallel mesh points computed, does 
not demand more memory bandwidth than what can be supplied by the FPGA's  
external DDR4 bandwidth. The FPGA's HBM memory can be used to support a larger 
$V$, which could then be limited by the resources available to implement the 
parallel compute pipelines. An estimate of maximum $V$ for an application can 
be computed by, using the FPGA operating frequency $f$, and maximum 
supported bandwidth of a data channel (or port) on the FPGA, $BW_{channel}$, 
and the size in bytes of a mesh element $sizeof(t)$ as follows:\vspace{-5pt}
\begin{equation}\label{eq/maxV}
BW_{channel} \geq 2Vf\times sizeof(t)\vspace{-0pt}
\end{equation}
For 2D meshes, if the width of the mesh $n$ is a multiple of vectorization 
factor V, then clock cycles for computing a single mesh point (or a cell) per 
iteration per compute module can be obtained from equation (\ref{eq/totclk2D}) as : \vspace{-5pt}
\begin{equation}\label{eq/clkcell2D}
Clks_{2D,cell} = 1/V + pD/2nV\vspace{-3pt}
\end{equation}
Setting $n$ to higher values gives a better clock cycles per mesh point 
ratio, the ideal being, $1/V$.  But higher order stencil applications on meshes 
with fewer rows will have a larger $(pD)/(2nV)$ value, indicating 
idling in the processing pipeline. We explore techniques to reduce this idle 
time in Section~\ref{subsec/batching}. 

A key parameter in (\ref{eq/totclk2D}) and (\ref{eq/totclk3D}) is the loop 
unroll factor, $p$ which directly determines performance, where a large $p$ 
reduces the total clock cycles required. However, $p$ is limited by the 
available resources on the FPGA as in Fig.~\ref{fig/unrolliter}, a larger $p$ 
requires more DSP blocks and LUTs. Furthermore, the internal memory required for a 
compute module, primarily due to memory capacity for the cyclic buffers also 
determines $p$. The number of DSP blocks required for a single mesh-point 
update, $G_{\mathit{dsp}}$ depends on the stencil loop kernel's arithmetic 
operations and number representation. Here we consider single precision floating point. With a $V$ vectorization factor, the total consumed is $V\times 
G_{\mathit{dsp}}$. If the total available DSP blocks on the FPGA is 
$\mathit{FPGA}_{\mathit{dsp}}$ then the maximum unroll factor based on DSP 
resources, $p_{\mathit{dsp}}$ is given by: \vspace{-3pt}
\begin{equation}\label{eq/pdsp}
{p_{\mathit{dsp}} = \mathit{FPGA}_{\mathit{dsp}}}/V 
G_{\mathit{dsp}}\vspace{-3pt}
\end{equation}
The internal memory requirement for a single compute module which performs a 
$D$ order stencil operation on an $m \times n$ mesh is $D\times m$. If the total 
available internal memory on the device is $\mathit{FPGA}_{\mathit{mem}}$, then 
maximum possible iterative unroll factor based on internal memory requirements, 
$p_{\mathit{mem}}$ is :\vspace{-3pt}
\begin{equation}\label{eq/pmem}
{p_{\mathit{mem}} = \mathit{FPGA}_{\mathit{mem}}}/{k D m}\vspace{-3pt}
\end{equation}
Here, $k$ is the size of a mesh element in bytes. The denominator of 
(\ref{eq/pmem}) becomes $kDmn$ for 3D meshes. Thus we see that the internal 
memory of an FPGA, directly limits the solvable mesh size. Usually, the 
above ideal depth is not achievable, as the FPGA internal memory, BRAMs and 
URAMs, are quantized (for example BRAMs are 18Kb/36Kb and URAMs are 288Kb 
on the U280). Additionally, the limited width configurations of the URAMs, plus 
the need to allow for flexible routing further reduces the effective internal 
memory resources. Thus we usually target an 80\%--90\% internal memory 
utilization. Then the maximum iterative loop unroll factor is given by the 
minimum of $p_{\mathit{dsp}}$ and $p_{\mathit{mem}}$. It is also worth 
considering that a larger pipeline depth, and hence more resource consumption 
leads to the design spreading over multiple SLRs. Communication between SLRs 
increasing routing congestion between these regions, directly impacting the 
achievable operating frequency.

\vspace{-0pt}
\section{Optimizations}\label{sec/opts}\vspace{-5pt}
\noindent Further optimizations and extensions are required to obtain high 
throughput for more complex applications. These include (1) spatial and temporal 
blocking, specifically for solvers over larger meshes, and (2) batching for 
improving performance and throughput of stencil applications on smaller meshes. 
In this section, we build on the baseline design from Section~\ref{sec/design} 
and extend the performance models to account for these optimizations.



\vspace{-3pt}
\subsection{Spatial and Temporal Blocking}\label{subsec/tiling}\vspace{-3pt}
\noindent The baseline design attempts to obtain perfect data reuse, requiring 
FPGA internal memory (consisting of BRAMs and URAMs) to be of size $D\times m$ 
for 2D and $D\times m\times n$ for 3D meshes. Equation (\ref{eq/pmem}) 
illustrates this, where the requirement becomes highly limiting for 
applications with higher order ($D$) stencils and/or on larger meshes 
(increasing $m$). Even if the mesh fully fits in the FPGA's DDR4 memory, a 
sufficiently large mesh could result in a $p_{\mathit{mem}}$ less than one, 
meaning that even a single compute module cannot be synthesized. A solution is 
to implement a form of spatial blocking, similar to cache blocking tiling on 
CPUs, for the FPGA. 


The idea is to use the baseline design to build an accelerator that 
operates on a smaller block of mesh elements and then transfer one such block at 
a time to the compute pipeline from FPGA DDR4 memory. The compute pipeline is 
designed with an appropriate vectorization factor ($V$) and an outer 
iterative loop unroll factor ($p$). Larger $p$ results in better exploitation 
of temporal locality, where the execution uses the same data several times. One 
issue with such a blocked execution is when applying the computation 
over the boundary of a block where a stencil computation on the boundary will not 
have the contributions from all the neighboring elements in the mesh. The 
solution is to overlap blocks such that the correct computation is carried out 
on the boundary by a subsequent block. The amount of overlap depends on the 
order of the stencil. Overlapping leads to redundant computation. However this 
overhead can be acceptable, due to the savings from further exploiting local 
data in multiple iterations. 

The main challenge of tiling then is to get close to maximum DDR4 memory 
bandwidth, due to the latency of smaller, non-contiguous  data transfer sizes.  
Such data transfers results due to a strided access pattern in one dimension 
when accessing memory locations within a spatial block. For example on the 
Xilinx U280, it takes 16 clock cycles to transfer 1024 Bytes via the 512 bit 
wide AXI interface bus, but the latency of the transfer is about 14 clock 
cycles. As such, multiple read/write requests should be made to hide the latency 
of each individual memory transaction. The preference to maintain a 512 bit wide 
bus interface to obtain better memory bandwidth further increases the amount of 
redundant computation at block boundaries as we must maintain a 512 bit 
alignment in read/write transactions, regardless of the order of the stencil. 


A final modification is the need to loop through the spatial blocks to solve 
over the full mesh. This control loop is best implemented on the FPGA to reduce 
latency due to the host calling multiple kernels on the FPGA. An important 
consideration is finding the optimal block size and its offset from the start of 
the mesh. The block size and offsets need only be computed once, which can be 
done on the host and copied to FPGA memory. Considering a 3D stencil application 
over a mesh of size $m\times n\times l$ solved by computing over with blocks (or 
tiles) of size $M\times N\times l$, the valid number of mesh points computed per 
block is given by: \vspace{-3pt}
\begin{equation}\label{eq/tiling_validpoints_perblock}
Block_{valid} = (M-pD)\times(N-pD)\times l \vspace{-3pt}
\end{equation}
Since the number of clock cycles required to process $p$ iterations (or a  
temporal block) on the $M\times N\times l$ spatial block is similar to 
the baseline design, the average time taken to compute one block (assuming 
block dimensions are a multiple of $V$) would be:\vspace{-5pt}
\begin{equation}\label{eq/tiling_clks_perblock}
Clks_{block, 3D} = \frac{M}{V}\times N \times \frac{l + pD/2}{p} \vspace{-3pt}
\end{equation}
Dividing (\ref{eq/tiling_validpoints_perblock}) by 
(\ref{eq/tiling_clks_perblock}) leads to the number of valid mesh points 
(or cells) computed per clock cycle (i.e. throughput, $T$) :\vspace{-3pt}
\begin{equation}\label{eq/tiling_cellsperclk}
T = (1 - \frac{pD}{M})\times(1 -\frac{pD}{N}) \times 
(\frac{pVl}{l + pD/2})\vspace{-3pt}
\end{equation}
Now, substituting $N$ from (\ref{eq/pmem}), for a 3D application, assuming 
full utilization of the FPGA's internal memory by a block, it can be shown that 
maximum throughput can be achieved for a given $p$ 
when :\vspace{-5pt}\begin{equation}\label{eq/tiling_maxthroughput_M}
M = \sqrt{FPGA_{mem}/kpD}\vspace{-3pt}
\end{equation}
The corresponding $N$, can be shown to be also equal to $M$, implying a square 
block to give the best throughput. However, the throughput also varies 
with $p$ and this can be analyzed by considering a square tile (i.e. $M=N$) 
applied to equation (\ref{eq/tiling_cellsperclk}) and assuming $l$ to be very 
large such that $\frac{l}{l + pD/2}$ is close to 1. With these assumptions, we 
can show that maximum throughput is achieved, for a given $M$, when setting $p$ 
to a $p_{\mathit{max}}$ given by:\vspace{-3pt}
\begin{equation}\label{eq/tiling_maxthroughput_p}
p_{\mathit{max}} = M/3D \vspace{-3pt}
\end{equation} 
Obtaining a value for $pV$ from (\ref{eq/pdsp}), assuming we use all the 
computational capacity of the FPGA, we can rewrite 
(\ref{eq/tiling_cellsperclk}) as: \vspace{-0pt}
\begin{equation}\label{eq/tiling_throughput_3D}
 T_{3D} = (1 - \frac{pD}{M})^2\times \frac{FPGA_{dsp}}{G_{dsp}}\times 
(\frac{l}{l + pD/2}) \vspace{-0pt}
\end{equation} 
The same for a 2D stencil application can also be derived as: \vspace{-0pt}
\begin{equation}\label{eq/tiling_throughput_2D}
 T_{2D} = (1 - \frac{pD}{M}) \times \frac{FPGA_{dsp}}{G_{dsp}}\times 
(\frac{n}{n + pD/2})\vspace{-0pt} 
\end{equation}
Here, we see that reducing pipeline depth $p$ and increasing $V$ will 
improve the performance of the spatial blocked design. The effect of $p$ is 
more significant for 3D applications. 

\vspace{-0pt}
\subsection{Batching}\label{subsec/batching}\vspace{-3pt}
\noindent A final optimization attempts to improve throughput for smaller mesh 
problems that usually perform poorly on accelerator platforms, including FPGAs. 
On traditional architectures such as GPUs the reason is the the 
under-utilization of the massive parallelism available. Essentially the time 
spent calling a kernel on the device and the overheads for data movement between 
host and device become dominate the actual processing time. 

On an FPGA, in addition to the above, further overheads are caused 
due to the latency of the processing pipeline, as given in equation 
(\ref{eq/clkcell2D}), compared to the time to process the mesh. The idle time is 
proportional to the width of the 2D mesh. Thus if a large number of smaller 
meshes are to be solved, as is the case in financial applications 
~\cite{RegulyBatching2019}, then processing one mesh at a time incurs 
significant latencies. This motivates the idea of grouping together meshes with 
the same dimensions in batches, increasing the overall throughput of the solve. 
In practice, the mesh can be extended in the last dimension by stacking up the 
small meshes. Now, the inter-compute module latencies only occur once at the 
start of the batched solve. With $B$, 2D meshes in a batch, the time to process 
a single mesh within a batched execution is given by: \vspace{-5pt} 
\begin{equation}\label{eq/clkmesh2D_batch}
Clks_{2D/batched\_mesh} = \left(\ceil*{\frac{m}{V}} \times (n + p\times\frac 
{D} {2B})\right)\vspace{-5pt} 
\end{equation}Thus, increasing $B$ significantly reduces the idle time from 
(\ref{eq/clkcell2D}). Similar reasoning can be applied for batched 3D 
meshes. 

\begin{table}[!t]\footnotesize
\centering
\caption{\small Experimental systems specifications.\normalsize}\vspace{-5pt}
\begin{tabular}{@{}ll@{}}
\toprule
FPGA	      & Xilinx Alveo U280~\cite{u280}                \\
\midrule
DSP blocks   & 8490 \\
BRAM / URAM   & 6.6MB (1487 blocks) / 34.5MB (960 blocks)\\
HBM           & 8GB, 460GB/s, 32 channels\\
DDR4          & 32GB, 38.4GB/s, in 2 banks (1 channel/bank)\\
Host	      & Intel Xeon Silver 4116 @2.10GHz (48 cores)\\
              & 256GB RAM, Ubuntu 18.04.3 LTS    \\
Design SW     & Vivado HLS, Vitis-2019.2 \\
\toprule
GPU	      & Nvidia Tesla V100 PCIe~\cite{u280}      \\
\midrule
Global Mem.   & 16GB HBM2, 900GB/s \\
Host	      & Intel Xeon Gold 6252 @2.10GHz (48 cores)\\
              & 256GB RAM, Ubuntu 18.04.3 LTS    \\
Compilers, OS  & nvcc CUDA 9.1.85, Debian 9.11 \\
\bottomrule
\end{tabular}\label{tab/systems}\vspace{-20pt}
\end{table}\normalsize

\begin{figure*}[t]\centering
\centering
\subfloat[][Baseline - 60000 iterations]
{\includegraphics[width=5.5cm]{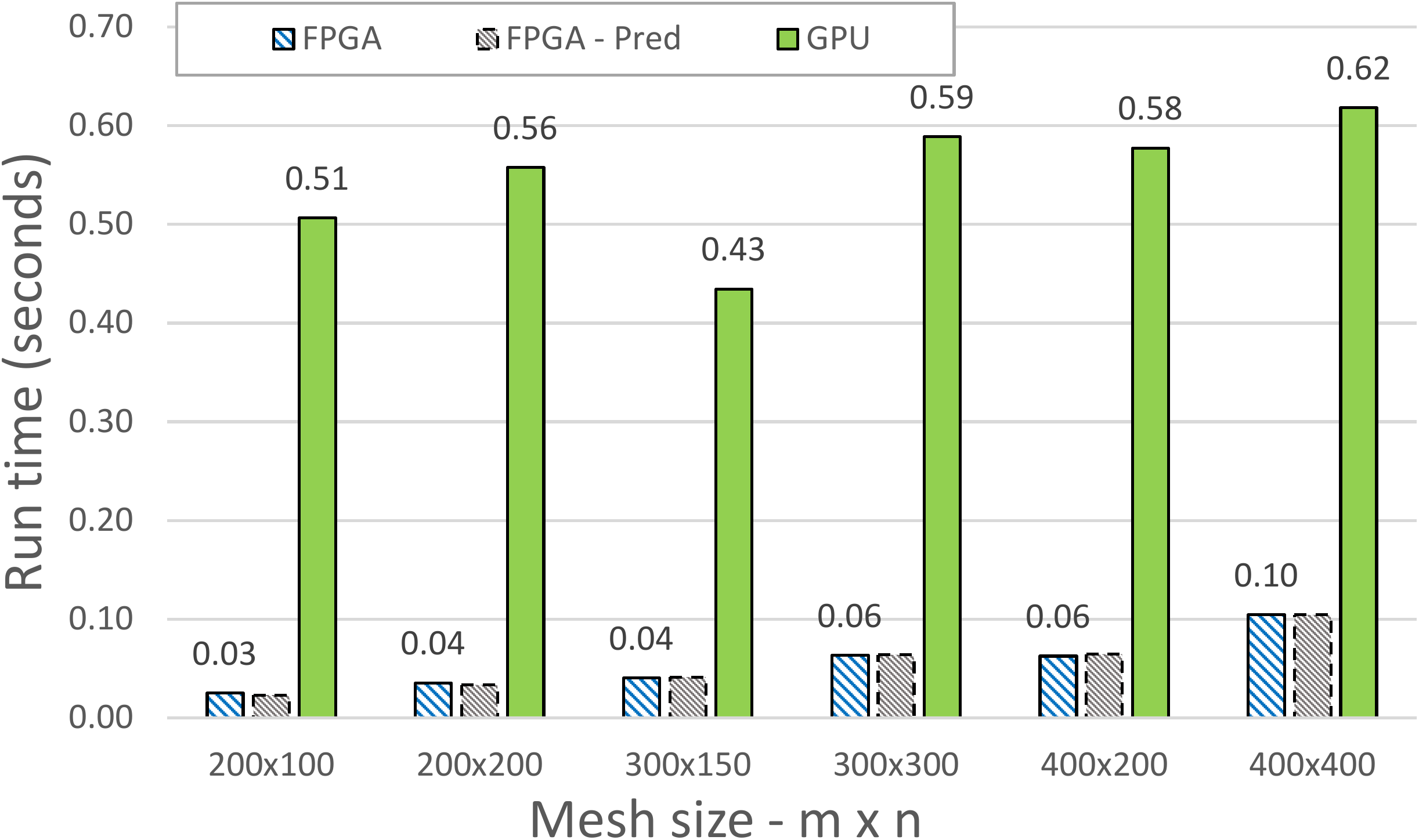}}
\subfloat[][Batching - 60000 iterations]
{\includegraphics[width=5.5cm]{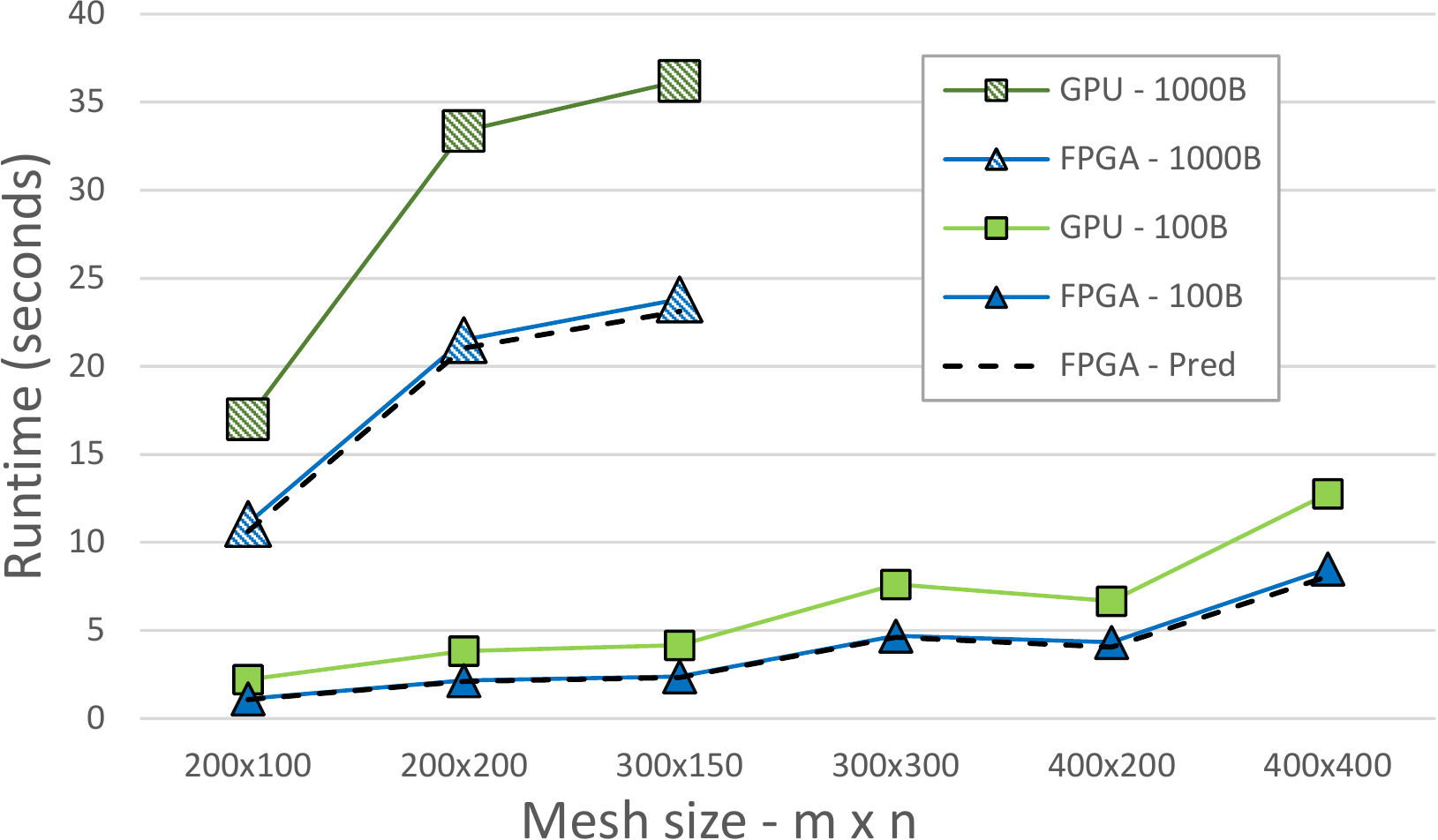}}
\subfloat[][Spatial-blocking - 6000 iterations]
{\includegraphics[width=5.5cm]{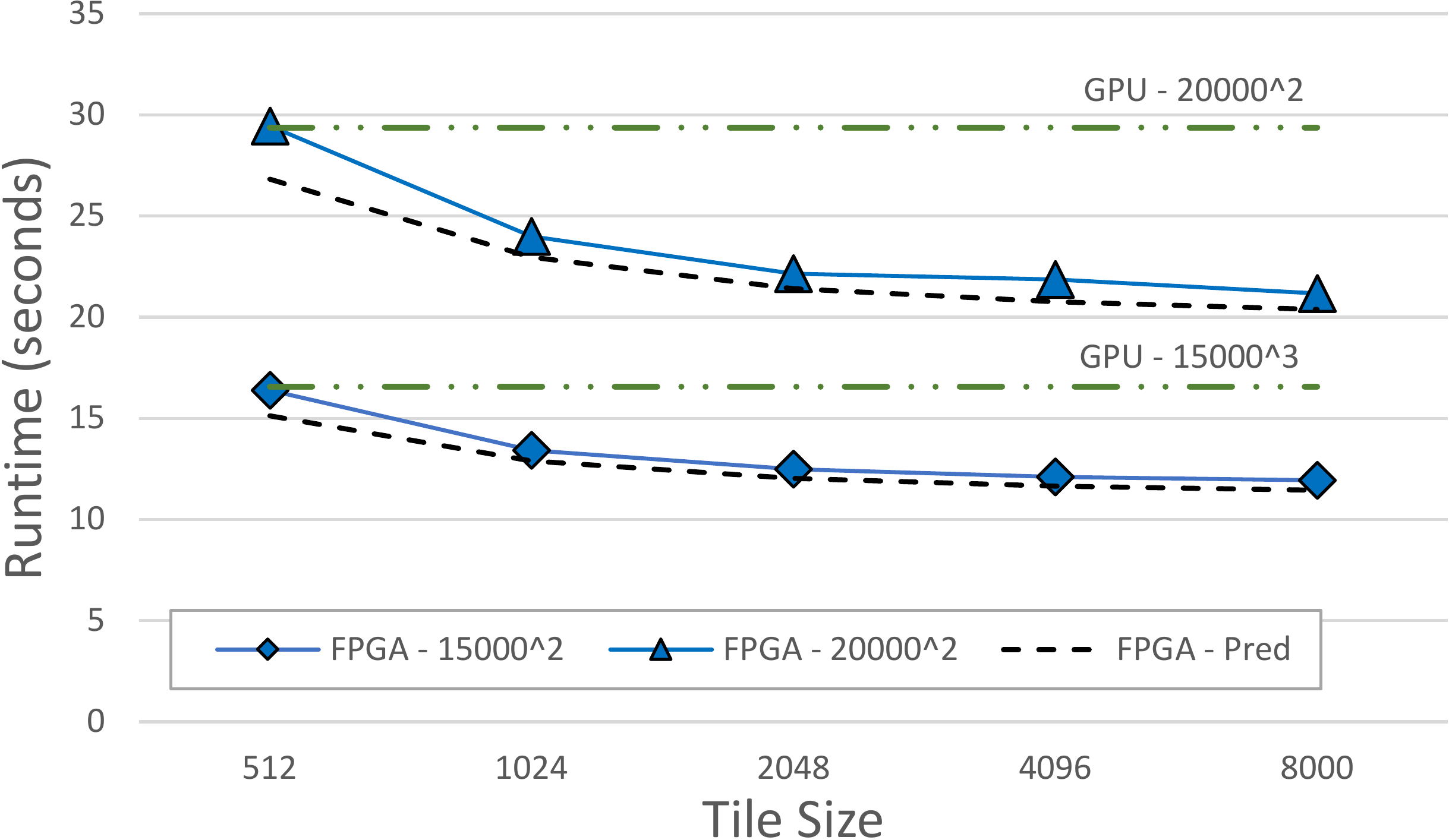}}\vspace{-5pt}
\caption{\small Poisson-5pt-2D 
performance}\vspace{-10pt}\label{grp/poisson_runtime}
\end{figure*}\vspace{-0pt}

\vspace{-0pt}\section{Performance}\label{sec/perf}\vspace{-3pt} 
\noindent In this section we apply the FPGA design strategy, optimizations, and 
extensions to illustrate their utility in accelerating stencil computations for 
explicit-iterative numerical solvers. We select three representative 
applications consisting of, both 2D and 3D, low and high order, and with single 
and multiple stencil loops to explore the versatility of our design flow. 
Model-predicted resource utilization estimates are used to determine initial 
design parameters, and runtime performance is compared to model predictions for 
each application. The implementations target the Xilinx Alveo U280 accelerator board 
and demonstrate concrete implementations for each application. We use Vivado 
C++ due to ease of use for configurations, arbitrary precision data types, and 
support of some C++ constructs compared to OpenCL, but note OpenCL can be 
equally used to implement the same design. Additionally, we compare equivalent 
implementations of each application's performance on a modern GPU 
system for comparison\footnote[1]{\noindent We have omitted CPU performance results here as our 
previous work~\cite{RegulyBatching2019} shows that GPUs provide significant 
speedups over CPUs for these applications}. 
\tablename{~\ref{tab/systems}} briefly details the specifications of the FPGA 
and GPU systems (both hardware and software) used in our experiments. 

\vspace{-3pt}
\subsection{Poisson-5pt-2D}\vspace{-3pt}
\noindent The first application is a 2D Poisson solver which uses a 
2nd order stencil, with scalar elements:\vspace{-5pt}
\begin{equation}\label{eq/poisson}
U^{t+1}_{i,j} = 
\tfrac{1}{8}\left(U^{t}_{i-1,j}+U^{t}_{i+1,j}+U^{t}_{i,j-1}+U^{t}_{i,j+1}
\right)+ \tfrac{1}{2}U^{t}_{i,j}\vspace{-5pt}
\end{equation}
A suitable initial vectorization factor $V$ can be identified by using 
(\ref{eq/maxV}) and assuming an operating frequency of 300MHz given this is 
the default set by the Vivado HLS tools. For a baseline implementation of 
Poisson a value of 8 for $V$ is calculated when using a single DDR4 channel 
or two HBM channels with a frequency of 300MHz. However, this frequency could 
only be supported when iterative loop unroll factor $p$ is in the order of 1--
20. Higher $p$ lead to routing congestion, which limited achievable frequency. As such the 
frequency was reduced to 250MHz to support a $p$ of 60, which we observed to 
give the best performance for this stencil. We find in some cases such a trial 
frequency adjustment is unavoidable, but our model significantly narrows the 
design space, enabling us to reason about and quickly obtain an optimal 
configuration. The number of DSP blocks required for a single mesh-point's 
stencil computation for Poisson and the resulting $p_{dsp}$ from (\ref{eq/pdsp}) 
for $V = 8$, assuming a 90\% DSP utilization, is given in the first row 
of~\tablename{ \ref{tab/modelparas_baseline_bathing}}.Column 4 gives the 
predicted $p_{dsp}$ from our performance model, while column 5 is the actual 
result after synthesis, indicating good agreement with the predicted design. 
\begin{table}[!t]\footnotesize
\centering 
\caption{\small Baseline and batching, model parameters}\vspace{-5pt}
\renewcommand{\arraystretch}{1.2}
\begin{tabular}{@{}lrrrr@{}}
\toprule
Application       & Freq. & $ G_{dsp}$ &  \multicolumn{2}{c}{$p_{dsp}$} 
\\
\cmidrule{4-5}
& (MHz) && (model) & (actual)\\
\midrule
Poisson-5pt-2D  & 250 & 14  &  68  &   60  \\
Jacobi-7pt-3D   & 246 & 33  &  28  &   29  \\
Reverse Time Migration             & 261 & 2444&  3  &   3  \\
\bottomrule
\end{tabular}\label{tab/modelparas_baseline_bathing}\normalsize\vspace{-10pt}
\end{table}
\begin{table}[t]\footnotesize
\centering 
\caption{\small Spatial blocking model parameters}\vspace{-5pt}
\renewcommand{\arraystretch}{1.2}
\begin{tabular}{@{}lrrrrrr@{}}
\toprule
App.            & $p$ &  $V$ &   $M$  & $N$ & $T_{2D|3D}$ & Valid ratio\\
\midrule
Poisson-5pt-2D  & 60  &  8   &   8192 &     & 472         & 98.5\% \\
Jacobi-7pt-3D   & 3  &  64   &   768  & 768 & 189       & 98.4\% \\\hline
\end{tabular}\label{tab/modelparas_tiling}\normalsize\vspace{-15pt}
\end{table}

\begin{figure*}[ht]\centering
\centering
\subfloat[][Baseline - 29000 iterations]
{\includegraphics[width=5.5cm]{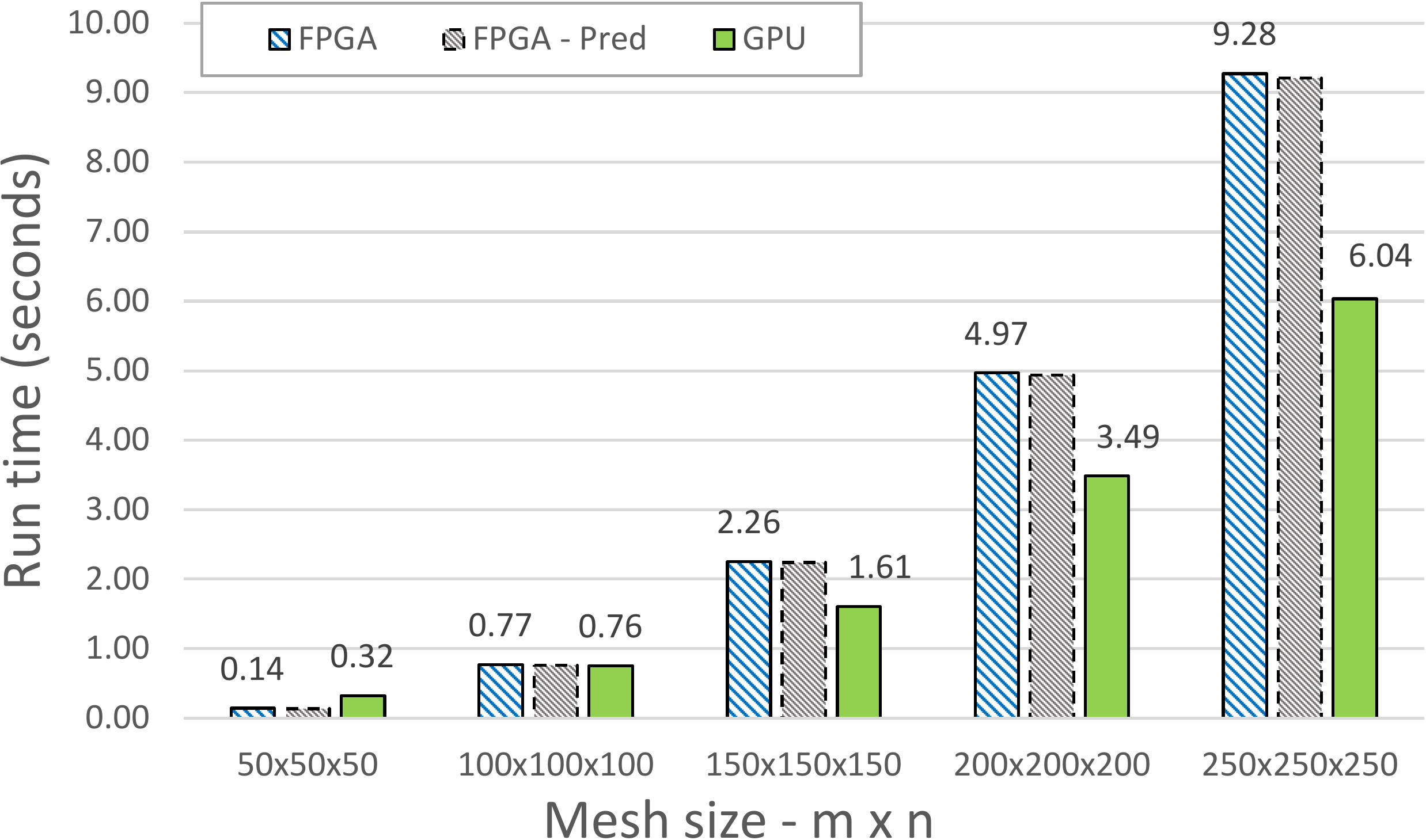}}
\subfloat[][Batching - 2900 iterations]
{\includegraphics[width=5.5cm]{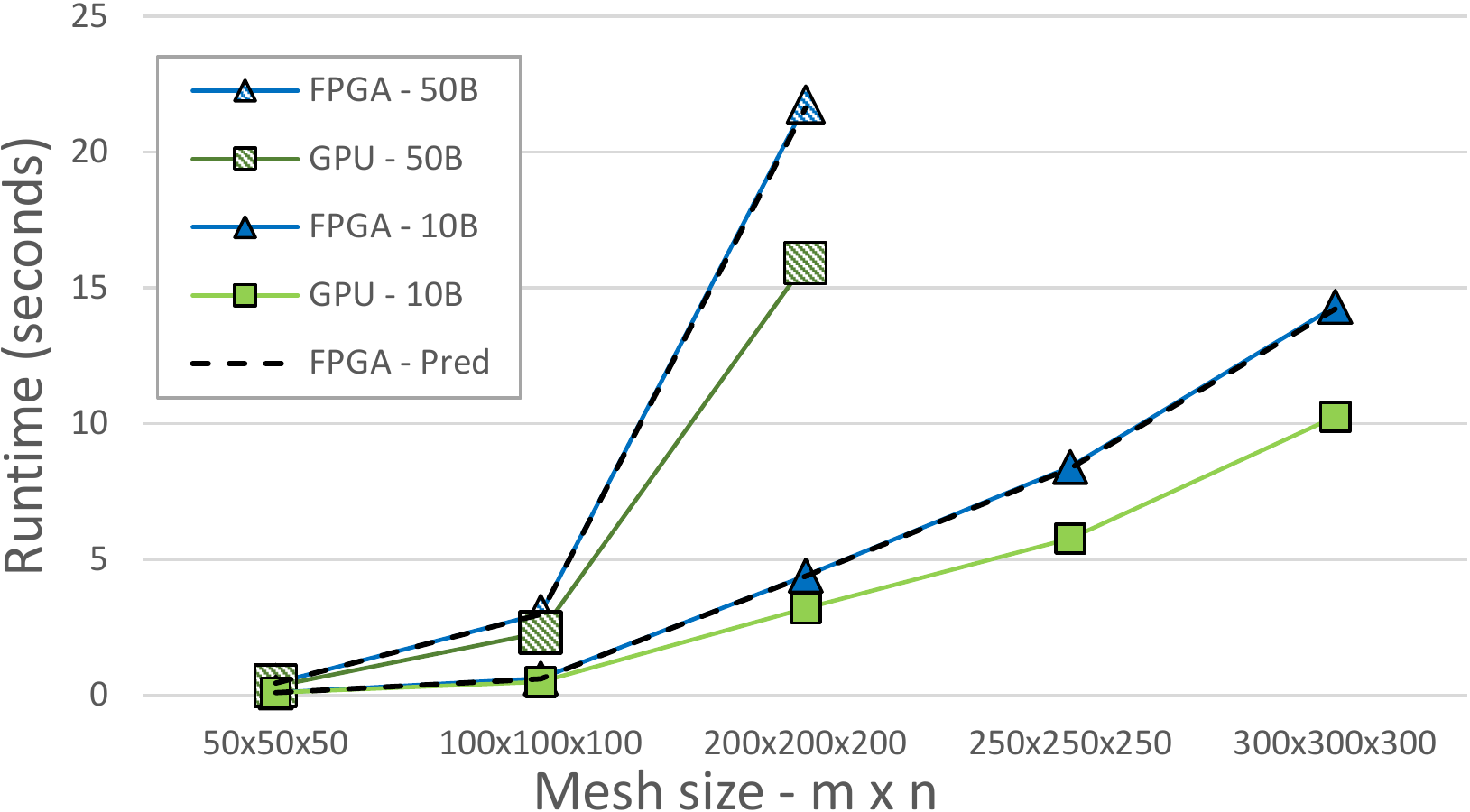}}
\subfloat[][Spatial-blocking - 120 iterations]
{\includegraphics[width=5.5cm]{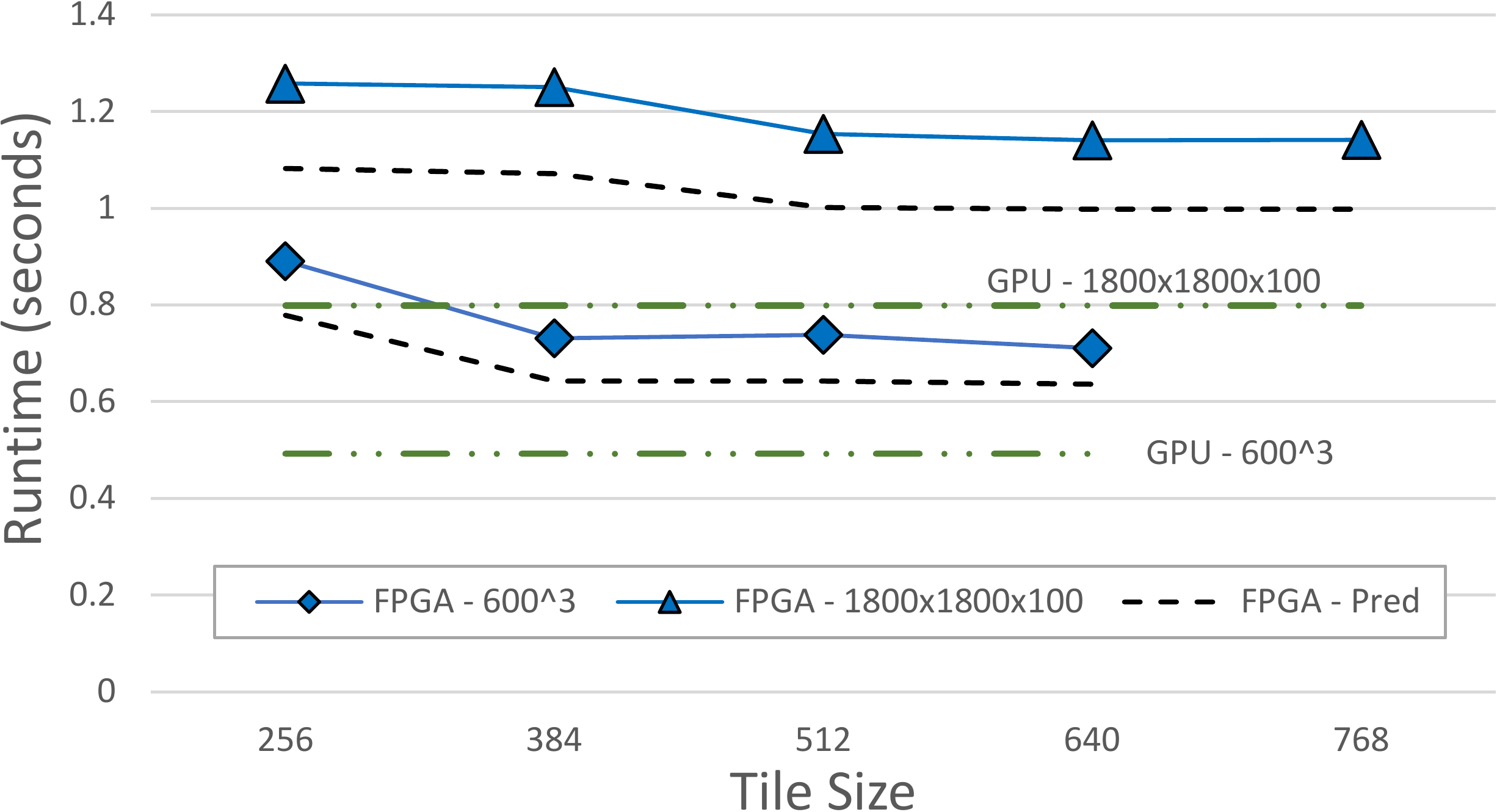}}\vspace{-5pt}
\caption{\small Jacobi-7pt-3D 
performance}\label{grp/Jac7pt3D_runtime}\vspace{-10pt}
\end{figure*}\vspace{-0pt}

\figurename{ \ref{grp/poisson_runtime}} (a) and (b) present the runtime 
performance of Poisson-5pt-2D, with the above design and compare the resultant 
performance to an equivalent implementation on the 
Nvidia V100 GPU. The achieved bandwidth and energy consumption from these 
runs are summarized in~\tablename{ \ref{tab/Poisson_BW_Energy}}. The bandwidth 
is computed by counting the total number of bytes transferred during the 
execution of the stencil loop (looking at the mesh data accessed) and dividing 
it by the total time taken by the loop. Baseline FPGA performance is 
significantly better than on the V100, since the GPU is not saturated by this 
application. The batching of 2D meshes as in~\cite{RegulyBatching2019} improves 
GPU performance significantly and offers a closer comparison. The FPGA achieves 
a maximum speedup of about 30--34\% for different mesh sizes and batching sizes 
of 100 (100B) and 1000 (1000B). Memory bandwidth results indicate high 
utilization of the communication channels in agreement with the observed 
runtimes. The \texttt{xbutil} utility was used to measure power during FPGA 
execution, while \texttt{nvidi-smi} was used for the same on the V100. The power 
consumption of the FPGA during the 1000B runs is indicative of the significant 
energy efficiency of the device compared to a GPU.  The FPGA was operating at an 
average 70W, while the GPU's power consumption ranged from 40W (for single 
batch) to 210W for 1000B runs on the larger meshes.\\
\begin{table}[t]\footnotesize
\centering \vspace{-5pt}
\caption{ \small Poisson-5pt: Bandwidth (GB/s) and Energy(kJ)}\vspace{-5pt}
\setlength\tabcolsep{3pt}
\begin{tabular}{@{}lrrrrrrrr@{}}
\toprule
\multicolumn{9}{c}{Baseline and Batching (60000 iterations)} \\
\midrule
Mesh  &  \multicolumn{2}{l}{Baseline} &   100B  &      & 1000B &     & 
\multicolumn{2}{l}{Energy-1000B}  \\
              &  FPGA & GPU &   FPGA  &  GPU & FPGA &GPU & FPGA & GPU 
\\
\midrule
$200\times100$  &  384  & 18  &   857 &  404 & 867 & 530 & 0.77 &3.48\\
$200\times200$  &  543  & 32  &   886 &  465 & 892 & 540 & 1.50 &6.74\\
$300\times150$  &  535  & 38  &   901 &  483 & 907 & 560 & 1.66 &7.60\\
$300\times300$  &  681  & 69  &   922   &  530 &     &     & & \\
$400\times200$  &  612  & 62  &   889   &  536 &     &     & & \\
$400\times400$  &  735  & 116 &   904   &  560 &     &     & & \\
\toprule
\end{tabular}\vspace{0pt}
\setlength\tabcolsep{4pt}
\begin{tabular}{@{}lrrrrr@{}}
\multicolumn{6}{c}{Spatial-blocking (60000 iterations)}\\
\toprule
Mesh&Tile Size & 
\multicolumn{2}{l}{BW} &\multicolumn{2}{l}{Energy}\\
&& FPGA & GPU & FPGA & GPU\\
\midrule
$15000^{2}$&1024&805&607&0.93&2.91\\
&4096&892&   &0.84& \\
&8000&905&   &0.83& \\
$20000^{2}$&1024&800&609& 1.67&4.96\\
&4096&879&   &1.52& \\
&8000&907&   &1.48& \\
\bottomrule
\end{tabular}\label{tab/Poisson_BW_Energy}\normalsize\vspace{-15pt}
\end{table}
\indent To implement Poisson-5pt-2D on larger meshes with spatial blocking, we 
assume a $V$ and $p$ equivalent to the baseline design and compute 
the valid mesh points updated per clock cycle using 
(\ref{eq/tiling_throughput_3D}). Here we assume the dimensions of the mesh to 
be very large. \tablename{ \ref{tab/modelparas_tiling}} lists the 
model parameters for spatial blocking. For Poisson we see that the 2D spatially 
blocked designs theoretically perform similar to the baseline design and thus 
we need not change the compute pipeline. Runtime, bandwidth and energy 
consumption of this implementation is given in \figurename{ 
\ref{grp/poisson_runtime}} (c) and~\tablename{ \ref{tab/Poisson_BW_Energy}}, 
respectively, including comparison to performance from the V100 GPU. Again we 
see good speedups and higher energy efficiency achieved with the FPGA, this 
time on large problem sizes with tiling.

\vspace{-0pt}
\subsection{Jacobi-7pt-3D}\vspace{-3pt}
\noindent The Jacobi iteration as a 3D, 7-point stencil, provides us 
with an initial, 3D, single stencil loop, for our evaluation:\vspace{-5pt}
\begin{equation}
\begin{aligned}U^{t+1}_{i,j,k} = &\ k_1U^{t}_{i+1,j,k}+ 
k_2U^{t}_{i-1,j,k}+ k_3U^{t}_{i,j-1,k}+ k_4U^{t}_{i,j,k}+ \\ 
& k_5U^{t}_{i,j+1,k}+ k_6U^{t}_{i,j,k+1}+ k_7U^{t}_{i,j,k-1} \ \ \ 
\ \ \ \ \ \ \ \ \ (18)\nonumber
\end{aligned}\vspace{-5pt}
\end{equation}
This application requires higher internal memory for the baseline design. For 
the spatially blocked design it involves transfers less than 4K from memory, 
which makes it difficult to approach raw external memory bandwidth. This is 
different to the baseline/batched and 2D spatially blocked design. We speculate 
that this could be the reason for the slightly less accurate model predictions 
in \figurename{ \ref{grp/Jac7pt3D_runtime}}(c). While the stencil is still 
fairly simple, now we see the GPU outperforming the FPGA conclusively, in both 
baseline, \figurename{ 
\ref{grp/Jac7pt3D_runtime}}(a) and batched \figurename{ 
\ref{grp/Jac7pt3D_runtime}}(b) tests. The V100 GPU gives nearly 40\% 
faster runtimes on the 50B problem. However, the FPGA remains 
more energy efficient for the same problem. For the $200\times 200$ problem with 50B, 
it is nearly 2$\times$ more energy efficient than the faster GPU run (see \tablename{ 
\ref{tab/Jac7pt3D_BW_Energy}}). The FPGA operated at an average 90W while 
the GPU power ranged from 77--240W. Spatial blocking was significantly more 
challenging and the resulting FPGA design, using a $640^2$  tile size was about 
40\% slower than the GPU runtime (see \figurename{ 
\ref{grp/Jac7pt3D_runtime}}(c)). However, the FPGA was again more energy 
efficient, operating at an average 70W consuming about 40--50\% less energy 
than the GPU (operating at 180--216 W) as seen in \tablename{ 
\ref{tab/Jac7pt3D_BW_Energy}}.

\begin{table}[t]\footnotesize
\centering \vspace{-5pt}
\caption{\small Jacobi-7pt-3D: Bandwidth (GB/s) and Energy(kJ)}\vspace{-5pt}
\setlength\tabcolsep{4pt}
\begin{tabular}{@{}lrrrrrrrr@{}}
\toprule
\multicolumn{9}{c}{Baseline (29k iters) and Batching (2.9k iters)} \\
\midrule
Mesh  &  \multicolumn{2}{l}{Baseline} &   10B  &      & 50B &     & 
\multicolumn{2}{l}{Energy-50B}  \\
        & FPGA& GPU & FPGA & GPU & FPGA& GPU & FPGA    & GPU \\
$50^3$  & 202 & 83  & 307  & 284 & 323 & 404 & 0.04    & 0.07 \\
$100^3$ & 301 & 284 & 378  & 434 & 387 & 469 & 0.27    & 0.51 \\
$200^3$ & 374 & 496 & 421  & 548 & 426 & 543 & 1.96    & 3.77 \\
$250^3$ & 391 & 559 & 431  & 585 &     &     &         & \\
$300^3$ & 403 & 553 & 438  & 569 &     &     &         & \\
\toprule
\end{tabular}\vspace{0pt}
\setlength\tabcolsep{4pt}
\begin{tabular}{@{}lrrrrr@{}}
\multicolumn{6}{c}{Spatial-blocking (120 iterations)}\\
\toprule
Mesh&Tile Size & 
\multicolumn{2}{l}{BW} & \multicolumn{2}{l}{Energy}\\
&& FPGA & GPU & FPGA & GPU\\
\midrule
$600^3$&256&233 &392& 0.062&0.106\\
&512 &281 &   & 0.051&   \\
&640  &292 &   & 0.049&   \\ 
$1800\times1800\times100$&256&247 &363& 0.088&0.143 \\
&512&270 &   & 0.080 &   \\
&640&273 &   & 0.079 &   \\
\bottomrule
\end{tabular}\label{tab/Jac7pt3D_BW_Energy}\normalsize\vspace{-15pt}
\end{table}


\begin{figure*}[ht]\centering
\centering
\subfloat[][Baseline - 1800 iterations]
{\includegraphics[width=7.5cm]{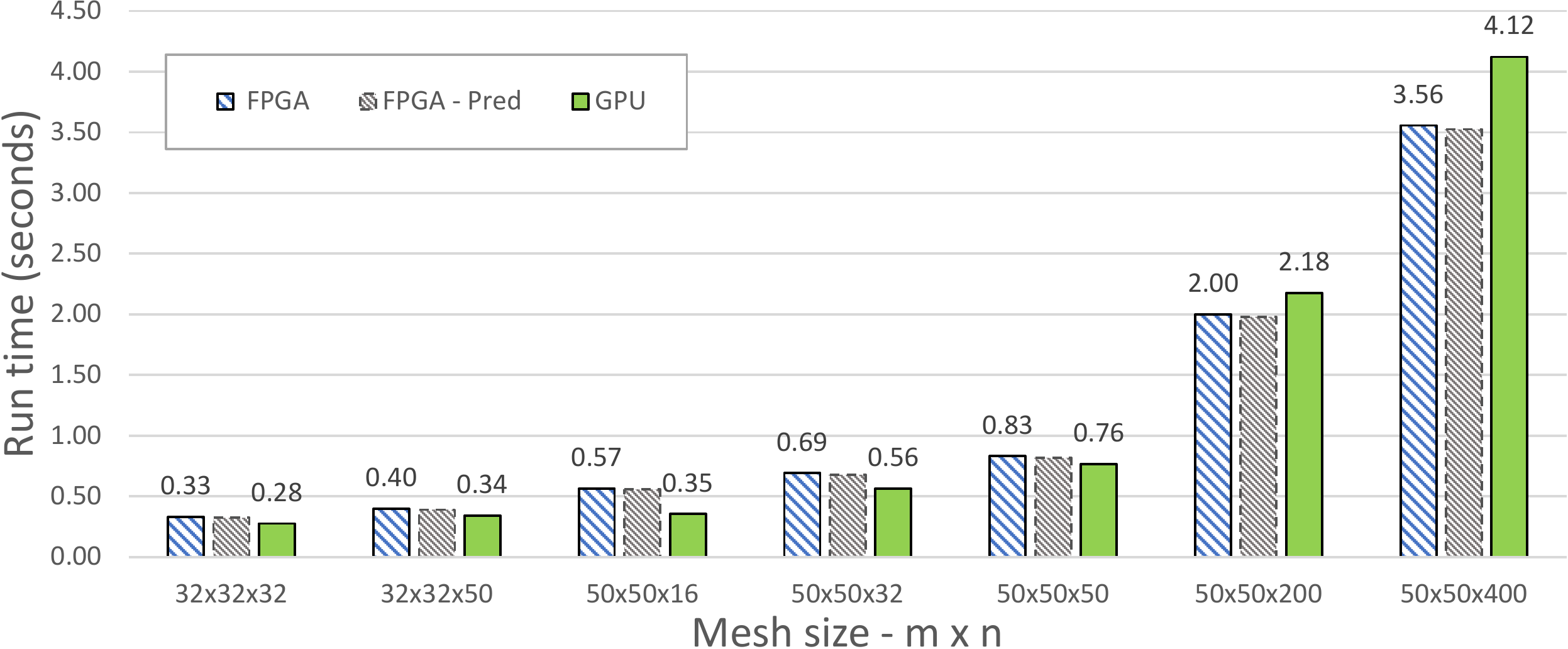}}\label{
grp/RTM_baseline}
\subfloat[][Batching - 180 iterations]
{\includegraphics[width=7.5cm]{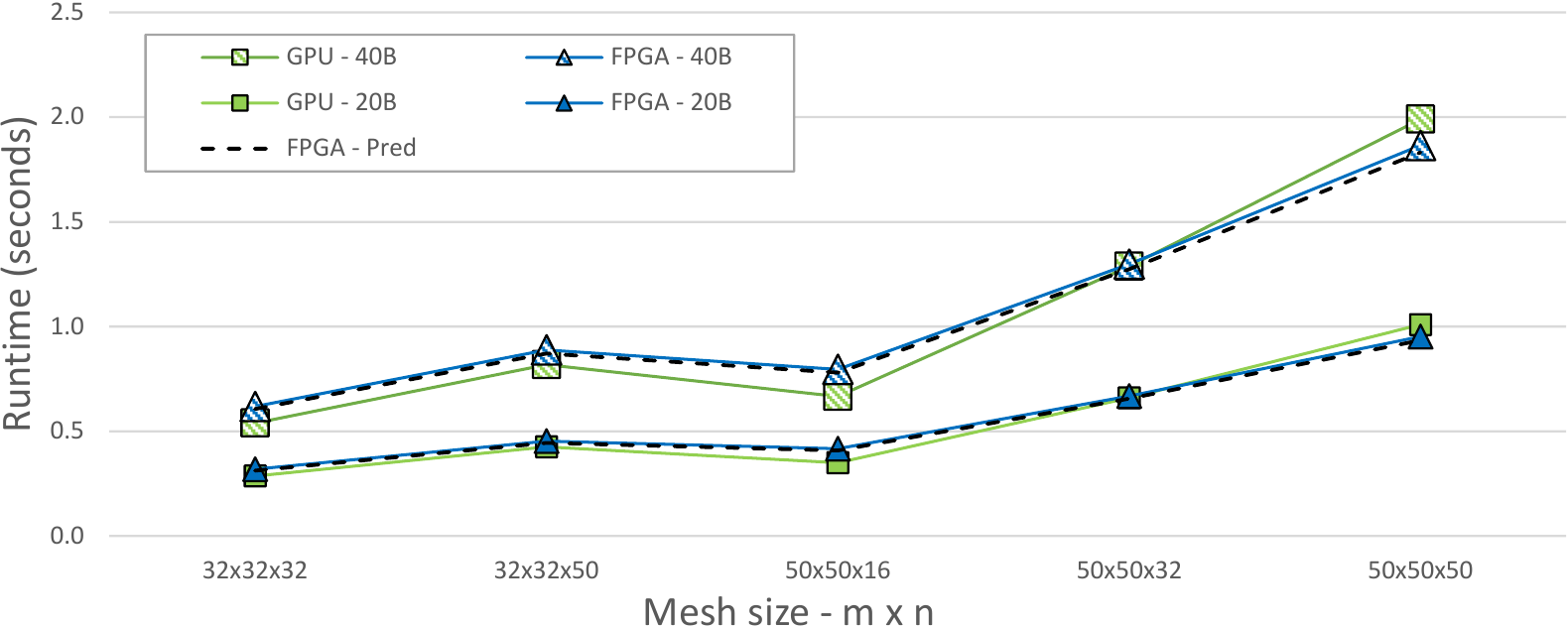} } \label {
grp/RTM_batched}\vspace{-5pt}
\caption{\small RTM performance}\label{grp/RTMBaseline_runtime}\vspace{-10pt}
\end{figure*}

\vspace{-3pt}
\subsection{Reverse Time Migration (RTM) - Forward Pass}\vspace{-3pt}
\noindent The final application we applied our development flow to is the 
forward pass from a Reverse Time Migration (RTM) solver~\cite{RTMApp}. The 
application represents algorithms of interest from industry~\cite{NAG}, going 
beyond simple single stencil loops. It includes an iterative loop consisting of 
multiple stencil loops as summarized in Algorithm~\ref{alg/RTM}. $Y, T$ and $K_1 
.. K_4$ are 3D floating-point (SP) data arrays defined on the mesh consisting of 
vector elements of size 6. $Y$ holds current values and $T$ holds intermediate 
values, both updated with the $f_{pml}$ function which uses a 25-point, eighth 
order 3D stencil. $K_1 .. K_4$ is accessed/updated with a self-stencil (or 
zeroth-order, i.e. $i,j,k$). $\rho$ and $\mu$ are two 3D scalar coefficient 
meshes, which are also accessed using a self-stencil. 
\begin{algorithm}[!h]\caption{\small RTM - Forward 
Pass}\label{alg/RTM}\small
\begin{algorithmic}
\For {$i=0, i< n_{iter}, i++$}\\
\ \ $K_{1} = f_{pml}(Y_{25pt}, \rho, \mu)\times dt;\ T = Y + K_{1}/2$ \\
\ \ $K_{2} = f_{pml}(T_{25pt}, \rho, \mu)\times dt;\ T = Y + K_{2}/2$ \\
\ \ $K_{3} = f_{pml}(T_{25pt}, \rho, \mu)\times dt;\ T = Y + K_{3}$ \\
\ \ $K_{4} = f_{pml}(T_{25pt}, \rho, \mu)\times dt$\\
\ \ $Y = Y + K_1/6 + K_2/3 + K_3/3 + K_4/6$
\EndFor
\end{algorithmic} 
\end{algorithm}\normalsize
\vspace{-5pt}
This application is significantly more complex than the previous applications 
and pushes the resource usage on the FPGA to its limits. Nevertheless our design 
strategy is able to provide a good implementation, albeit limited to a batched 
design. The number of stencil loops was reduced by fusing the $K_1,K_2,$ and 
$K_3$ with the corresponding $T$ loop. The $K_4$ and final $Y$ update were 
merged into one further loop, resulting in a total of 4 loops. For an FPGA 
implementation, all the four fused loops needed to be brought into a single 
pipeline. Intermediate data $T$ and $K_1 ... K_4$ were replaced with a FIFO 
stream connected through window buffers. Similarly $\rho, \mu$ and $Y$ were 
internally buffered and fed to subsequent compute units. These optimizations 
reduce the number of memory accesses to a single read and write of $Y$ and a 
single read each for $\rho$ and $\mu$. These are significant savings compared to 
the original loop chain.

A limitation of the FPGA implementation is that the mesh plane 
size (in this 3D application), is limited to $64^2$ as it uses 3D stencils on a 
6 dimensional element (i.e. a vector of 6 floats). Furthermore, partitioning 
four compute-intensive kernels on the U280's three SLR regions was a 
significant challenge. Our implementation avoids spanning of a compute unit on 
multiple SLRs to avoid inter SLR routing congestion, by setting $V$ to 1, 
allowing us to fit the four fused loops in one SLR. This, then allows for an 
iterative loop unroll factor of 3 ($p$) given the three SLRs on the U280. We do 
note that using more HBM channels could provide more bandwidth to obtain a 
larger $V$, but we have not explored this in current work. A solution for the 
limited mesh size is of course spatial blocking, but it requires $p = 4$. This 
leads to a tile size dimension $M = 96$ from (\ref{eq/tiling_maxthroughput_p}) 
given $D$ is 8, which requires a large amount of FPGA internal memory, making an 
implementation on the U280 challenging as the four fused loops will span across 
SRLs. We leave this to future work. 

\begin{table}[!h]\footnotesize
\centering \vspace{-5pt}
\caption{ \small RTM - Avg. Bandwidth (GB/s) and Energy(kJ)}\vspace{-5pt}
\renewcommand{\arraystretch}{1.2}
\setlength\tabcolsep{3pt}
\begin{tabular}{@{}lrrrrrrrr@{}}\midrule
\multicolumn{9}{l}{Baseline (1800 iters) and Batching (180 iters)} \\
\toprule
Mesh  &  \multicolumn{2}{l}{Baseline} &     \multicolumn{2}{l}{20B}  &   \multicolumn{2}{l}{40B}& 
\multicolumn{2}{l}{Energy-40B}  \\
      &  FPGA     & GPU &   FPGA  &  GPU & FPGA  & GPU & FPGA    & GPU \\
      \midrule
$32\times32\times32$ &  108 & 130 & 225 & 251 & 232 & 266 & 0.043 & 0.086 \\
$32\times32\times50$ &  141 & 163 & 247 & 263 & 253 & 274 & 0.062 & 0.133 \\
$50\times50\times16$ &  77  & 124 & 210 & 251 & 220 & 263 & 0.055 & 0.111 \\
$50\times50\times32$ &  127 & 155 & 262 & 266 & 270 & 272 & 0.091 & 0.218 \\
$50\times50\times50$ &  165 & 179 & 287 & 271 & 293 & 275 & 0.130 & 0.338 \\
\bottomrule
\end{tabular}\label{tab/RTMB_BW_Energy}\normalsize\vspace{-5pt}
\end{table}

\noindent From the runtime results in \figurename{ 
\ref{grp/RTMBaseline_runtime}} and bandwidth results in \tablename{ 
\ref{tab/RTMB_BW_Energy}} we see that the FPGA implementation is either matching 
or marginally better performing than the GPU. Note that, given there are 
four stencil loops fused on the FPGA the bandwidth reported is for the fused 
loop. The GPU bandwidth therefore is the average for the full loop chain. The 
most time consuming kernel, $f_{pml}$ on the GPU achieved around 180 GB/s, 
while the highest bandwidth achieved by a single kernel is over 340GB/s. Again 
we see that the FPGA operates at a lower average power (70W) than the 
GPU (51--170W) consuming 2$\times$ less energy.







\vspace{-0pt}
\section{Conclusions}\label{sec/conclusions}\vspace{-3pt}
\noindent In this paper we developed a unified workflow and a supporting 
predictive analytic model for FPGA synthesis of structured-mesh stencil 
applications that combines standard state-of-the-art techniques with a number of 
high-gain optimizations targeting features of real-world work loads. The model 
allows estimation of design parameters, resource usage, and performance for performant FPGA implementation. The workflow was 
applied to three representative applications, implemented on a 
Xilinx Alveo U280 FPGA. Performance was compared to highly-optimized HPC-grade Nvidia V100 GPU code. In most cases, the FPGA is able to match 
or improve on GPU performance. However, even when runtime is inferior to the GPU, significant energy savings, over 2$\times$ for the 
largest application, are observed. Estimations produced by the model were shown to be accurate and a good guide in the design process. Future work will investigate how 
a similar workflow can be applied to implicit solvers and further automating the 
development of this class of application on FPGAs, including alternative numerical representations. The FPGA and GPU source code
developed in this paper are available 
at~\cite{KKrepo}.\small\vspace{0pt}\normalsize






\vspace{-5pt}
\section*{Acknowledgment}\vspace{-3pt} \small
\noindent Gihan Mudalige was supported by the Royal Society Industry Fellowship 
Scheme(INF/R1/1800 12). Istv\'an Reguly was supported by National Research, Development and  Innovation 
Fund of Hungary, project PD 124905, financed under the PD\_17 funding scheme. The authors would like to thank Jacques Du Toit and Tim
Schmielau at NAG UK Ltd., for the RTM application and for their valuable 
advice. \normalsize\vspace{-5pt}

\bibliographystyle{IEEEtran}
\bibliography{Bib}\vspace{-10pt}

\begin{thebibliography}{10}
\providecommand{\url}[1]{#1}
\csname url@samestyle\endcsname
\providecommand{\newblock}{\relax}
\providecommand{\bibinfo}[2]{#2}
\providecommand{\BIBentrySTDinterwordspacing}{\spaceskip=0pt\relax}
\providecommand{\BIBentryALTinterwordstretchfactor}{4}
\providecommand{\BIBentryALTinterwordspacing}{\spaceskip=\fontdimen2\font plus
\BIBentryALTinterwordstretchfactor\fontdimen3\font minus
  \fontdimen4\font\relax}
\providecommand{\BIBforeignlanguage}[2]{{%
\expandafter\ifx\csname l@#1\endcsname\relax
\typeout{** WARNING: IEEEtran.bst: No hyphenation pattern has been}%
\typeout{** loaded for the language `#1'. Using the pattern for}%
\typeout{** the default language instead.}%
\else
\language=\csname l@#1\endcsname
\fi
#2}}
\providecommand{\BIBdecl}{\relax}
\BIBdecl

\bibitem{cousins2016designing}
D.~B. Cousins, K.~Rohloff, and D.~Sumorok, ``{Designing an FPGA-accelerated
  homomorphic encryption co-processor},'' \emph{IEEE Transactions on Emerging
  Topics in Computing}, vol.~5, no.~2, pp. 193--206, 2016.

\bibitem{owaida2017centaur}
M.~Owaida, D.~Sidler, K.~Kara, and G.~Alonso, ``{Centaur: A framework for
  hybrid CPU-FPGA databases},'' in \emph{Proceedings of the IEEE International
  Symposium on Field-Programmable Custom Computing Machines (FCCM)}, 2017, pp.
  211--218.

\bibitem{wang2016dlau}
C.~Wang, L.~Gong, Q.~Yu, X.~Li, Y.~Xie, and X.~Zhou, ``{DLAU: A scalable deep
  learning accelerator unit on FPGA},'' \emph{IEEE Transactions on
  Computer-Aided Design of Integrated Circuits and Systems}, vol.~36, no.~3,
  pp. 513--517, 2016.

\bibitem{becker2015maxeler}
T.~Becker, O.~Mencer, S.~Weston, and G.~Gaydadjiev, ``{Maxeler data-flow in
  computational finance},'' in \emph{FPGA Based Accelerators for Financial
  Applications}, 2015, pp. 243--266.

\bibitem{fahmy2015virtualized}
S.~A. Fahmy, K.~Vipin, and S.~Shreejith, ``{Virtualized FPGA accelerators for
  efficient cloud computing},'' in \emph{Proceedings of the IEEE International
  Conference on Cloud Computing Technology and Science (CloudCom)}, 2015, pp.
  430--435.

\bibitem{Reguly2017}
I.~Z. Reguly, G.~R. Mudalige, and M.~B. Giles, ``{Loop tiling in large-scale
  stencil codes at run-time with OPS},'' \emph{IEEE Transactions on Parallel
  and Distributed Systems}, vol.~29, no.~4, pp. 873--886, April 2018.

\bibitem{InPar2012}
\BIBentryALTinterwordspacing
G.~R. Mudalige, M.~B. Giles, I.~Z. Reguly, C.~Bertolli, and P.~H.~J. Kelly,
  ``{OP2: An active library framework for solving unstructured mesh-based
  applications on multi-core and many-core architectures},'' \emph{2012
  Innovative Parallel Computing, InPar 2012}, 2012. [Online]. Available:
  \url{http://dx.doi.org/10.1109/InPar.2012.6339594}
\BIBentrySTDinterwordspacing

\bibitem{pyfr2016}
P.~Vincent, F.~Witherden, B.~Vermeire, J.~S. Park, and A.~Iyer, ``{Towards
  green aviation with python at petascale},'' in \emph{SC16: International
  Conference for High Performance Computing, Networking, Storage and Analysis},
  Nov 2016, pp. 1--11.

\bibitem{devito2018}
\BIBentryALTinterwordspacing
F.~{Luporini}, M.~{Lange}, M.~{Louboutin}, N.~{Kukreja}, J.~{H{\"u}ckelheim},
  C.~{Yount}, P.~{Witte}, P.~H.~J. {Kelly}, F.~J. {Herrmann}, and G.~J.
  {Gorman}, ``{Architecture and performance of Devito, a system for automated
  stencil computation},'' \emph{CoRR}, vol. abs/1807.03032, Jul 2018. [Online].
  Available: \url{http://arxiv.org/abs/1807.03032}
\BIBentrySTDinterwordspacing

\bibitem{3drtm_2011}
\BIBentryALTinterwordspacing
H.~Fu and R.~G. Clapp, ``{Eliminating the memory bottleneck: An FPGA-based
  solution for 3D reverse time migration},'' in \emph{Proceedings of the 19th
  ACM/SIGDA International Symposium on Field Programmable Gate Arrays}, ser.
  FPGA ’11.\hskip 1em plus 0.5em minus 0.4em\relax New York, NY, USA:
  Association for Computing Machinery, 2011, p. 65–74. [Online]. Available:
  \url{https://doi.org/10.1145/1950413.1950429}
\BIBentrySTDinterwordspacing

\bibitem{SDSLc2015}
\BIBentryALTinterwordspacing
P.~Rawat, M.~Kong, T.~Henretty, J.~Holewinski, K.~Stock, L.-N. Pouchet,
  J.~Ramanujam, A.~Rountev, and P.~Sadayappan, ``{SDSLc: A multi-target
  domain-specific compiler for stencil computations},'' in \emph{Proceedings of
  the 5th International Workshop on Domain-Specific Languages and High-Level
  Frameworks for High Performance Computing}, ser. WOLFHPC ’15.\hskip 1em
  plus 0.5em minus 0.4em\relax New York, NY, USA: Association for Computing
  Machinery, 2015. [Online]. Available:
  \url{https://doi.org/10.1145/2830018.2830025}
\BIBentrySTDinterwordspacing

\bibitem{Waidyasooriya2017}
H.~M. {Waidyasooriya}, Y.~{Takei}, S.~{Tatsumi}, and M.~{Hariyama},
  ``{OpenCL-based FPGA-platform for stencil computation and its optimization
  methodology},'' \emph{IEEE Transactions on Parallel and Distributed Systems},
  vol.~28, no.~5, pp. 1390--1402, 2017.

\bibitem{soda2018}
Y.~{Chi}, J.~{Cong}, P.~{Wei}, and P.~{Zhou}, ``{SODA: Stencil with pptimized
  dataflow architecture},'' in \emph{2018 IEEE/ACM International Conference on
  Computer-Aided Design (ICCAD)}, 2018, pp. 1--8.

\bibitem{Zohouri2018}
\BIBentryALTinterwordspacing
H.~R. Zohouri, A.~Podobas, and S.~Matsuoka, ``{Combined spatial and temporal
  blocking for high-performance stencil computation on FPGAs using OpenCL},''
  in \emph{Proceedings of the 2018 ACM/SIGDA International Symposium on
  Field-Programmable Gate Arrays}, ser. FPGA ’18.\hskip 1em plus 0.5em minus
  0.4em\relax New York, NY, USA: Association for Computing Machinery, 2018, p.
  153–162. [Online]. Available: \url{https://doi.org/10.1145/3174243.3174248}
\BIBentrySTDinterwordspacing

\bibitem{Zohouri2018a}
H.~R. {Zohouri}, A.~{Podobas}, and S.~{Matsuoka}, ``{High-performance
  high-order stencil computation on FPGAs using OpenCL},'' in \emph{2018 IEEE
  International Parallel and Distributed Processing Symposium Workshops
  (IPDPSW)}, 2018, pp. 123--130.

\bibitem{Waidyasooriya2019}
H.~M. {Waidyasooriya} and M.~{Hariyama}, ``{Multi-FPGA accelerator architecture
  for stencil computation exploiting spacial and temporal scalability},''
  \emph{IEEE Access}, vol.~7, pp. 53\,188--53\,201, 2019.

\bibitem{HeteroCL2019}
\BIBentryALTinterwordspacing
Y.-H. Lai, Y.~Chi, Y.~Hu, J.~Wang, C.~H. Yu, Y.~Zhou, J.~Cong, and Z.~Zhang,
  ``{HeteroCL: A multi-paradigm programming infrastructure for software-defined
  reconfigurable computing},'' in \emph{Proceedings of the 2019 ACM/SIGDA
  International Symposium on Field-Programmable Gate Arrays}, ser. FPGA
  ’19.\hskip 1em plus 0.5em minus 0.4em\relax New York, NY, USA: Association
  for Computing Machinery, 2019, p. 242–251. [Online]. Available:
  \url{https://doi.org/10.1145/3289602.3293910}
\BIBentrySTDinterwordspacing

\bibitem{Dohi2013}
K.~{Dohi}, K.~{Fukumoto}, Y.~{Shibata}, and K.~{Oguri}, ``{Performance modeling
  and optimization of 3D stencil computation on a stream-based FPGA
  accelerator}, year={2013}, pages={1-6},'' in \emph{2013 International
  Conference on Reconfigurable Computing and FPGAs (ReConFig)}.

\bibitem{Natale2016}
G.~{Natale}, G.~{Stramondo}, P.~{Bressana}, R.~{Cattaneo}, D.~{Sciuto}, and
  M.~D. {Santambrogio}, ``{A polyhedral model-based framework for dataflow
  implementation on FPGA devices of iterative stencil loops},'' in \emph{2016
  IEEE/ACM International Conference on Computer-Aided Design (ICCAD)}, 2016,
  pp. 1--8.

\bibitem{DSLSC_2013}
K.~{Sano}, Y.~{Hatsuda}, and S.~{Yamamoto}, ``{Multi-FPGA accelerator for
  scalable stencil computation with constant memory bandwidth},'' \emph{IEEE
  Transactions on Parallel and Distributed Systems}, vol.~25, no.~3, pp.
  695--705, 2014.

\bibitem{3dverilog2009}
M.~{Shafiq}, M.~{Pericàs}, R.~{de la Cruz}, M.~{Araya-Polo}, N.~{Navarro}, and
  E.~{Ayguadé}, ``{Exploiting memory customization in FPGA for 3D stencil
  computations},'' in \emph{2009 International Conference on Field-Programmable
  Technology}, 2009, pp. 38--45.

\bibitem{vhdltemplate2012}
M.~{Schmidt}, M.~{Reichenbach}, and D.~{Fey}, ``{A generic VHDL template for 2D
  stencil code applications on FPGAs},'' in \emph{2012 IEEE 15th International
  Symposium on Object/Component/Service-Oriented Real-Time Distributed
  Computing Workshops}, 2012, pp. 180--187.

\bibitem{SSA2011}
K.~{Sano}, Y.~{Hatsuda}, and S.~{Yamamoto}, ``{Scalable streaming-array of
  simple soft-processors for stencil computations with constant
  memory-bandwidth},'' in \emph{2011 IEEE 19th Annual International Symposium
  on Field-Programmable Custom Computing Machines}, 2011, pp. 234--241.

\bibitem{licht2020stencilflow}
\BIBentryALTinterwordspacing
J.~de~Fine~Licht, A.~Kuster, T.~D. Matteis, T.~Ben-Nun, D.~Hofer, and
  T.~Hoefler, ``{StencilFlow: mapping large stencil programs to distributed
  spatial computing systems},'' 2020. [Online]. Available:
  \url{https://arxiv.org/abs/2010.15218}
\BIBentrySTDinterwordspacing

\bibitem{XilinxSSI2012}
``Xilinx - large fpga methodology guide,'' 2012, \url{
  https://www.xilinx.com/support/documentation/sw_manuals/xilinx14_7/ug872_largefp
  ga.pdf}.

\bibitem{RegulyBatching2019}
I.~Z. Reguly, B.~Moore, T.~Schmielau, J.~du~Toit, and G.~R. Mudalige, ``{Batch
  solution of small PDEs with the OPS DSL},'' in \emph{High Performance
  Computing}, M.~Weiland, G.~Juckeland, S.~Alam, and H.~Jagode, Eds.\hskip 1em
  plus 0.5em minus 0.4em\relax Cham: Springer International Publishing, 2019,
  pp. 124--141.

\bibitem{u280}
\emph{Alveo U280 data center accelerator card data sheet}, Xilinx Inc., May
  2020, v1.3.

\bibitem{RTMApp}
R.~Clayton and B.~Engquist, ``{Absorbing boundary conditions for acoustic and
  elastic wave equations},'' \emph{Bulletin of the Seismological Society of
  America}, vol.~67, no.~6, pp. 1529--1540, 12 1977.

\bibitem{NAG}
``Discussions with the {Numerical Algorithms Group, UK.}'' 2019-2020.

\bibitem{KKrepo}
``{High-Level FPGA accelerator design for structured-mesh-based explicit
  numerical solvers - GitHub Code Repository},'' 2020,
  \url{https://github.com/Kamalavasan/StencilsOnFPGA}.

\end{thebibliography}





\end{document}